\documentclass[paper,longauth,traditabstract]{aa} 
\usepackage{graphicx}
\usepackage[varg]{txfonts}
\usepackage{natbib,twoopt}
\usepackage{epstopdf}
\usepackage{slashbox}
\usepackage{verbatim}

\usepackage{longtable}
\usepackage{rotating}
\usepackage{pdflscape}
\usepackage{multirow}
\usepackage{amstext}
\usepackage{color}

\begin{document}

\def\HII{H\,{\sc{ii}}\,}
\def\mm{\,$\mu$m\,}
\def\spitzer{$\it{Spitzer}$\,}
\def\herschel{$\it{Herschel}$\,}

\title{
Spatial distribution of star formation related to\\
 ionized regions throughout the inner Galactic plane
}

   \author{
     P. Palmeirim\inst{1,2}
     \and
     A. Zavagno\inst{1}   
     \and  
D. Elia\inst{3}
\and
T. J. T. Moore\inst{4}
\and
A. Whitworth\inst{5}
\and
P. Tremblin\inst{6}
\and
A. Traficante\inst{3}
\and
M. Merello\inst{3}
\and
D. Russeil\inst{1}
\and
S. Pezzuto\inst{3}
\and
L. Cambr\'esy\inst{7}
\and
A. Baldeschi\inst{3}
\and
M. Bandieramonte\inst{8}
\and
U. Becciani\inst{9}
\and
M. Benedettini\inst{3}
\and
C. Buemi\inst{9}
\and
F. Bufano\inst{9}
\and
A. Bulpitt\inst{10}
\and
R. Butora\inst{11}
\and
D. Carey\inst{10}
\and
A. Costa\inst{9}
\and
L. Deharveng\inst{1}
\and
A. Di Giorgio\inst{3}
\and
D. Eden \inst{4}
\and
A. Hajnal\inst{12}
\and
M. Hoare\inst{10}
\and
P. Kacsuk\inst{12}
\and
P. Leto\inst{9}
\and
K. Marsh\inst{5}
\and
P. M\`{e}ge\inst{1}
\and
S. Molinari\inst{3}
\and
M. Molinaro\inst{11}
\and
A. Noriega-Crespo\inst{13}
\and
E. Schisano\inst{3}
\and
E. Sciacca\inst{9}
\and
C. Trigilio\inst{9}
\and
G. Umana\inst{9}
\and
F. Vitello\inst{9}
           }

   \institute{
         Aix Marseille Univ, CNRS, LAM, Laboratoire d'Astrophysique de Marseille, Marseille, France 
        \email{pedro.palmeirim@lam.fr}
        \and
        Instituto de Astrof\'isica e Ci{\^e}ncias do Espa\c{c}o, Universidade do Porto,
        CAUP, Rua das Estrelas, PT4150-762 Porto, Portugal \\
        \email{pedro.palmeirim@astro.up.pt}
        \and
        Istituto di Astrofisica e Planetologia Spaziali INAF-IAPS, via Fosso del Cavaliere 100, I-00133 Roma, Italy
        \and
        Astrophysics Research Institute, Liverpool John Moores University, Liverpool Science Park Ic2, 146 Brownlow Hill, Liverpool L3 5RF, UK
        \and
        School of Physics \& Astronomy, Cardiff University, Cardiff, CF24, 3AA, UK
        \and
        Maison de la Simulation, CEA-CNRS-UPS-UVSQ, USR 3441, Centre d'étude de Saclay, 91191 Gif- Sur-Yvette, France
        \and
        Universit\'e de Strasbourg, CNRS, Observatoire astronomique de Strasbourg, UMR 7550, F-67000 Strasbourg, France
        \and
        CERN, Route de Meyrin 385, 1217 Meyrin , Switzerland
        \and
        INAF-Osservatorio Astrofisico di Catania, Via S. Sofia 78, 95123 Catania, Italy
        \and
       School of Physics and Astronomy, University of Leeds, Leeds, UK
        \and    
        INAF - Osservatorio Astronomico di Trieste, via G.B. Tiepolo 11, Trieste, Italy
        \and
        MTA-SZTAKI, 1111 Budapest, Kende u. 13-17, Hungary
        \and
        Space Telescope Science Institue, 3700 San Martin Drive, Balitmore MD, USA
                 }
   \date{}
   
 \abstract{
We present a comprehensive statistical analysis of star-forming objects located in the vicinities of
1~360 bubble structures throughout the Galactic Plane and their local environments.
The compilation of $\sim$70~000 star-forming sources,
found in the proximity of the ionized (H{\sc{ii}}) regions and
detected in both Hi-GAL and GLIMPSE surveys,
 provided a broad
 overview of the different evolutionary stages of star-formation in bubbles, from prestellar objects to more evolved young stellar objects (YSOs).
Surface density maps of star-forming objects clearly reveal an 
evolutionary trend where more evolved star-forming objects (Class II YSO candidates) are found spatially located near the center,
while younger star-forming objects are found 
at the edge of the bubbles.
We derived dynamic ages for a subsample of 182 \HII regions for 
which kinematic distances and radio continuum flux measurements 
were available.
We detect approximately 80\% more star-forming sources per unit area in the direction of bubbles
than in the surrounding fields. 
We estimate the clump formation efficiency (CFE) of Hi-GAL clumps in the direction of the shell of the bubbles to be $\sim 10\%$,
around twice the value of the CFE in fields that are not affected by feedback effects.
We find that the higher values of CFE are mostly due to the higher 
CFE of protostellar clumps, in particular in younger bubbles,
whose density of the bubble shells is higher.
We argue that the formation rate from prestellar to protostellar phase 
is probably higher during the early stages of the (\HII) bubble expansion.
Furthermore, we also find a higher fraction of massive YSOs (MYSOs) in bubbles 
at the early stages of expansion ($< 2$~Myr) than older bubbles.
Evaluation of the fragmentation time inside the shell of bubbles
advocates the preexistence of clumps in the medium 
before the bubble expansion
in order to explain the formation of MYSOs in the youngest \HII regions ($< 1$~Myr), 
as supported by numerical simulations.
Approximately 23\% of the Hi-GAL clumps are found located in the direction of a bubble, with 15\% for prestellar clumps
and 41\% for protostellar clumps. We argue that the high fraction of protostellar clumps may be due to
the acceleration of the star-formation process cause by the feedback of the (H{\sc{ii}}) bubbles.
}
  

   \maketitle
%
\section{Introduction}

Thanks to ground- and space-based surveys of the Galactic Plane,
 a new picture of the interstellar medium has emerged. 
In particular, {\it Spitzer} images 
at 8 $\mu$m and 24 $\mu$m 
revealed an almost ubiquitous presence of  
bubble structures seen throughout the entire Galactic Plane \citep{Churchwell2006,Churchwell2007}. 
\citet{Anderson2011} have found that half of the \HII regions have a bubble morphology.
\HII regions are
generated by massive stars that ionize the surrounding medium, 
causing it to expand isotropically.
The 8 $\mu$m emission appears in the {\it Spitzer} images as a ring-like structure
caused by the emission of polycyclic aromatic hydrocarbon (PAH) molecules
that are concentrated in the shell of the bubble, which are excited by the UV emission of the ionizing source(s) 
 \citep{Leger1984,Drain2007,Tielens2008}.
On the other hand, the 24~$\mu$m emission is typically seen 
to radiate in the inner part of the bubbles, but its origin is still
 unclear \citep[see][]{Deharveng2010}. 
It is normally attributed 
 to very small dust grains that can survive at shorter distances from the ionizing sources \citep{Desert1990},
 but some evidence also suggests that 24~$\mu$m emission may be caused by large dust grains in thermal equilibrium
  \citep{Cesarsky2000}.

As bubbles expand, they interact with the surrounding cold molecular medium, 
which may trigger star formation.
Three main physical mechanisms are 
considered when the triggering of star-formation is discussed:
the collect and collapse (C\&C) \citep{Elmegreen1977}, the radiation-driven implosion (RDI) \citep[][]{Deharveng2010},
and the enhancement of preexisting density substructures and subsequent global implosion (EDGI) \citep{Walch2015}.
The C\&C mechanism causes the cold gas and dust 
 to be swept up by the supersonic expansion of an \HII region and
to be compressed in between the ionization and shock fronts.
As a result, the compressed collected material can reach high densities,
 become gravitationally unstable, and fragment into massive dense clumps \citep{Whitworth1994,Deharveng2005,Zavagno2010}.
Conversely, the RDI mechanism arises from 
 the interaction of the ionizing radiation with the turbulent surrounding medium.
This process is responsible for the triggering of preexisting condensations of dense gas, 
 as well as for the formation of pillar structures \citep{Hester1996,Tremblin2012}.
The EDGI mechanism
describes the expansion of the \HII regions in a fractal medium,
taking the preexistence of dense structures into account, 
which through a hybrid interaction of the C\&C and RDI processes 
 trigger on much shorter timescales ($< 1$~Myr),
as demonstrated in numerical simulations \citep{Walch2012,Walch2013,Walch2015}.

Star formation is indeed observed at the edges of ionized H\,{\sc{ii}} regions, as shown by 
several observational studies \citep{Zavagno2007,Zavagno2010,Deharveng2010,Samal2014,Liu2016}.
In particular, \citet{Thompson2012} and \citet{Kendrew2012} made a detailed statistical 
analysis using {\it Spitzer} bubbles, which revealed
that massive young stellar objects (YSOs) 
identified in the Red MSX Source survey \citep{Urquhart2008}
were typically observed at the edge of the bubbles,
while intrinsically red sources from the \citet{Robitaille2008} catalog 
exhibit a flat distribution inside the bubbles.
Since the \citet{Robitaille2008} sample contained more evolved objects,
Thompson et al. suggested that this may be due to an evolutionary gradient 
across the bubble.
Moreover, about $\sim 25\%$ of MYSOs 
throughout the Galactic Plane are located
at the edges of \HII regions,
which means that 
around one quarter of MYSOs formed in the Milky Way 
might have been triggered by \HII regions.

Although several results suggest that triggered star formation may be occurring,
 the exact relation between the presence of an \HII region, its expansion, 
 and star formation observed in its vicinity is difficult to establish \citep{Dale2015}. 
Nevertheless, the link between \HII regions and triggered star-formation
 may be better addressed by studying several triggering indicators 
in a large sample of \HII regions.

In the present work, we aim to exploit the combined information of new-generation surveys of the Galactic Plane to deliver 
a global understanding of the star formation in \HII regions and their local environments
for a large sample of 1360 bubbles that are located at various
places throughout the Galactic Plane.
In particular, \herschel data from the 
\herschel Infrared Galactic plane survey \citep[Hi-GAL;][]{Molinari2010,Molinari2016}
allow us to detect star-forming objects 
in earlier evolutionary stages and
view the cold dust in their surroundings. 
Combined with the detection of
YSO at a more advance evolutionary stage
using the \spitzer GLIMPSE source catalog, we wish to 
obtain a broad overview of
the different evolutionary stages of star formation in bubbles.

This paper is organized as follows. In Sect.~\ref{sec:data} we describe the observations and their respective data reduction procedures, and 
present the sample of (H{\sc{ii}}) bubbles we used that are located in the inner Galactic plane.
Section~\ref{sec:sources} presents the sample selection and classification of GLIMPSE YSOs and Hi-GAL clumps found in the vicinities of (H{\sc{ii}})  bubbles.
Section~\ref{sec:res} presents the results, which are discussed in Sect.~\ref{sec:disc}. Summary and conclusions are presented in Sect.~\ref{sec:conc}.  

\section{Observations and data reduction}\label{sec:data}

\subsection{{\it Herschel} Hi-GAL observations}

We made use of \herschel observations \citep{Pilbratt2010} of the inner part of the Galactic Plane ($68^{\circ} \ge l \ge -70^{\circ}$ and $|b| \le 1^{\circ}$), which were obtained as part of the 
Herschel Infrared Galactic plane survey \citep[Hi-GAL;][]{Molinari2010,Molinari2016}.
The Galactic Plane was covered by individual observations of $\sim 2^{\circ}.2$ tiles
that were mapped  in two orthogonal scan directions at $60'' s^{-1}$, from which we
simultaneously obtained PACS \citep{Poglitsch2010}
 70~$\mu$m and 160~$\mu$m and SPIRE \citep{Griffin2010} at 250~$\mu$m, 350~$\mu$m, and  500~$\mu$m
images using the parallel mode of \herschel.

The data reduction was performed using the ROMAGAL data-processing code for both PACS and SPIRE 
(see Traficante et al. 2011 for details)\nocite{Traficante2011}.
The raw data were first processed up to level 0.5 using standard steps in the pipeline of the \herschel Interactive Processing Environment (HIPE) \citep{Ott2010}.
Post-processing was carried out with a dedicated pipeline written by the Hi-GAL Consortium
(see Molinari et al. 2016 for details),\nocite{Molinari2016}
which removes glitches and slow thermal drifts.
The final PACS and SPIRE maps were produced by a map-making algorithm
based on generalized least squares (GLS) \citep{Natoli2001}.

Zero-level offsets were added to the \herschel images based on the comparison between the \herschel data with IRIS (Improved Reprocessing of the IRAS Survey) all-sky maps
and Planck data at comparable wavelengths \citep[cf.][]{Bernard2010,Molinari2016}.
 
\subsection{Galactic Plane bubble sample}

For this work we made use of 
the {\it The Milky Way Project} (MWP\footnote{http://www.milkywayproject.org}) catalog \citep{Simpson2012}. 
This catalog was obtained by compressing a very large sample of visual measurements on the {\it Spitzer}-GLIMPSE 8~$\mu$m and 24~$\mu$m maps, 
performed by  over 35~000 independent volunteers 
that are part of the large community of the ``citizen scientists''  Zooniverse\footnote{http://zooinverse.org} project. 
The MWP bubbles were classified into two separate categories, small and large catalogs, based on their sizes.
Since the MWP small bubble catalog contains bubbles that are not well resolved in the {\it Spitzer} images, 
we opted to select bubbles from the large catalog,
 which provides more reliable statistical measurements.
 
To ensure that the selected MWP bubbles were resolved in the \herschel maps, we
opted to only select from the catalog bubbles with 
effective radius\footnote{$R_{\rm eff}$ is defined in the MWP catalog as the mean between the geometrically averaged semi-minor and semi-major axis of the inner and outer ellipses of the bubble.}
($R_{\rm eff}$) larger than 72$\arcsec$ 
(approximately twice the full-width at half maximum ($FWHM$) of the {\it Herschel} SPIRE 500~$\mu$m beam). 
This led to a total sample of 1360 bubbles and covers the $|l| < 65^{\circ}$ and $|b| < 1^{\circ}$ range of the Galactic Plane.

 \section{Source detection and classification}\label{sec:sources}

 \subsection{Hi-GAL prestellar and protostellar objects}\label{sec:higal}

To probe recent star formation activity in the vicinities of bubbles,
we made use of the physical Hi-GAL source catalog (Elia et al. 2017)\nocite{Elia2017}.
The Hi-GAL sources were first extracted from the Hi-GAL PACS and SPIRE images using the CuTeX photometry code \citep{Molinari2010},
from which we generated an individual photometric catalog for each band for the inner Galactic plane, as described in \citet{Molinari2016}.
The single-band catalogs were then compiled into a band-merged catalog following \citet{Elia2010,Elia2013}.
A final sample of 100~922 sources
 with detection in at least three consecutive bands between 160 $\mu$m and 500 $\mu$m were considered
eligible for the fit of a modified blackbody to their spectral energy distributions (SEDs) 
and were included in the Hi-GAL physical catalog, 
with their respective derived physical properties.
Moreover, sources were classified according to their evolutionary stages following the same approach as described in \citet{Elia2013}.
A 70~$\mu$m counterpart was used to distinguish protostellar and starless sources, and the latter
were further classified 
into gravitationally unbound and bound (prestellar) sources, based on the mass threshold inferred by Larson's third law.
Potential biases due to issues such as distances, multiplicities, and saturations are addressed in Elia et al. (2017).

For the present work we selected from the catalog all prestellar (starless and gravitationally bound) and protostellar (with a 70 $\mu$m counterpart) 
objects that were found within four times the effective radius ($<4~R_{\rm eff}$)
of a bubble. This led to
a total sample of 25~911 prestellar and 14~918 protostellar 
sources that were used in our study.
Since the large majority of the detected sources in the Galactic Plane are located at distances between $2-15$~kpc (see text bellow),
we expect their physical sizes to be greater then 0.1 pc. 
Thus, we henceforth define them as clumps \citep[e.g.,][]{Bergin2007}.

However, we would like to note that considering detection at 70~$\mu$m as a selection criteria for the protostellar clumps located in the vicinity of \HII regions can cast some doubt on their reliability. The bulk of the emission of \HII regions is caused by the photodissociation region (PDR) formed at the edges that strongly emit at 70~$\mu$m \citep{Paladini2012}. Owing to the nebular morphology of the PDR, it may be possible to find some compact knots that can led to a false classification of a protostellar clump
that can lead to the classification of a prestellar
clump as a protostellar clump
\citep[e.g.,][]{Gaczkowski2013}. To address this issue and the reliability of our protostellar clump sample, we used the information of mid-infrared (MIR) counterparts
that is available in the Hi-GAL catalog, since protostellar sources also emit in the MIR, but the effect from the PDR is considerably fainter at this wavelength.
For each Hi-GAL source a MIR counterpart was searched for at 24, 22, and 21~$\mu$m in the respective MIPSGAL \citep{Gutermuth2015}, WISE \citep{Wright2010}, and MSX \citep{Egan2003} data, when available (see Elia et al. 2017 for further details)\nocite{Elia2017}. Using this information, we find that $\sim90~\%$ of our protostellar clump sample has at least one counterpart measurement at one of the aforementioned wavelengths in the MIR. Therefore, the level of potential contamination that is due to PDR is not considered in the further analysis.

\subsection{GLIMPSE YSO source classification}\label{sec:ysos}

As protostellar objects accrete most of their surrounding envelope, 
they evolve into a YSO,
at which point outflows start to further clear the envelope,
shifting the peak emission toward mid-infrared wavelengths.
\spitzer, with its optimized wavelength coverage to
detect and characterize the different evolutionary stages of YSOs,
has been used extensively in previous works \citep[e.g.,][]{Allen2004,Evans2009,Heiderman2010,Rebull2010}.

In this work, the initial selection of YSO candidates was based on the color criteria of 
IRAC photometric measurements made available
in the GLIMPSE Source Catalog (I + II  + 3D) \citep{Glimpse2009},
following a similar approach as 
\citet{Gutermuth2008}.
Only sources with IRAC magnitudes brighter than 14.2, 14.1, 11.9, and 10.8 mag at 3.6, 4.5, 5.8, and 8.0 $\mu$m 
were considered to ensure 
a detection reliability above 98\% in all IRAC bands,
 based on the incompleteness levels 
 inferred in the 
GLIMPSE~I Assurance Quality Document\footnote{http://www.astro.wisc.edu/glimpse/GQA-master.pdf}. 
In order to preserve sources with good-quality photometry, we 
avoided detections with an adjacent companion 
by verifying that their close source flag
(csf)\footnote{Flag sources in the Archive Catalog with a companion within 3$\arcsec$.} 
was equal to zero and that the magnitude errors were lower than 0.2 mag \citep{Robitaille2008}.
 
    \begin{figure}
       \begin{minipage}{1\linewidth}
     \resizebox{\hsize}{!}{
   \includegraphics[angle=0]{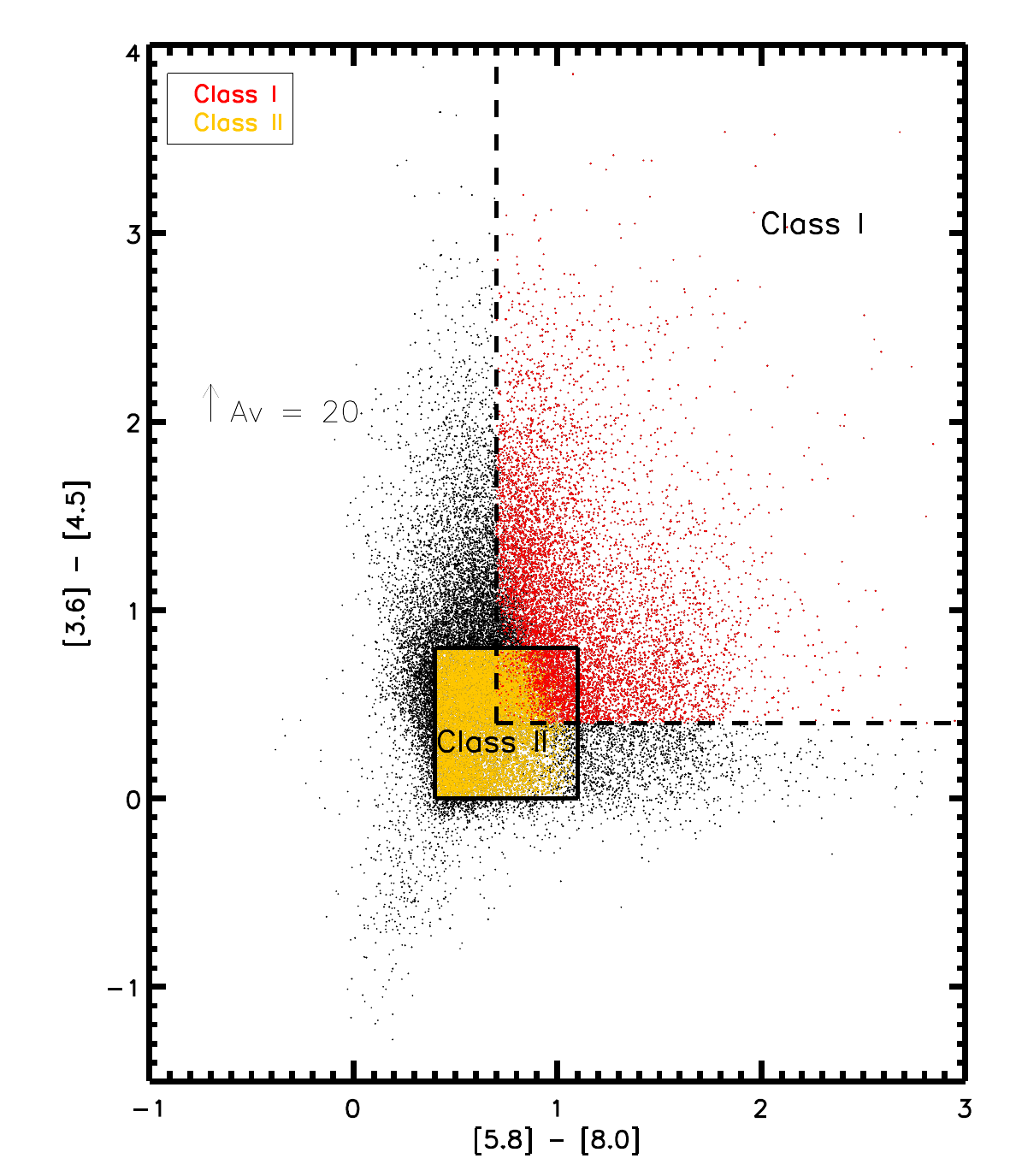}
   }
    \end{minipage}   
   \caption{IRAC color-color diagram of 91~02{\bf6} YSO candidates located at less than four times the effective radius ($R_{\rm eff}$) of 
   the 1360 bubbles.
   The solid and dashed lines represent the respective boundaries of Class I and II sources, following \citet{Allen2004}.
  The red and yellow dots represent the Class I (10~694) and Class II (18~209) YSO candidates
  selected based on both their spectral indices and position on the diagram, respectively. 
  The reddening vector of $A_{\rm V}$ = 20 mag was derived from the extinction law of \citet{Indebetouw2005}.
   }
              \label{irac}
   \end{figure}

For the selection criteria of YSO candidates we followed the
\citet{Gutermuth2009} complementary analysis of the Bootes field of IRAC data \citep{Stern2005},
which allows removing potential contamination of star-forming galaxies and active galactic nuclei (AGNs). 
In practice, 
YSO candidates were selected by applying the following color conditions:
$$[4.5] - [8.0] > 0.5,$$ 
$$[3.6] - [5.8] > 0.35,$$ 
$$[3.6] - [5.8] \le 3.5 \times ([4.5] - [8.0]) - 1.25.$$

These selection criteria were applied to all GLIMPSE sources 
found within $4 \times R_{\rm eff}$ of a bubble, which
led to a total of 91~026 YSO candidates.

The selected YSO candidates were further classified into different evolutionary stages according
 to their infrared spectral index $\alpha_{\rm IRAC} = d~{\rm log}(\nu F_{\nu})/d~{\rm log}(\nu)$, as defined by \citet{Lada1987},
 using all four IRAC fluxes. 
The spectral index $\alpha_{\rm IRAC}$ was determined from the slope of the SED 
measured between 3.6 $-$ 8.0 $\mu$m
 \citep[see][]{Lada2006}. 
The flux error measurements were used as the quadratic weight when performing the least-squares fits. 
YSO candidates with $\alpha_{\rm IRAC} > -0.3$ were classified as Class I and $-0.3 > \alpha_{\rm IRAC} > -1.6$ as Class II{\bf.}
Since Class III candidates ($-1.6 > \alpha_{\rm IRAC} > -2.56$) are the most affected by biases,
they were discarded from our analysis.
The main reason for this exclusion is the high level of contamination of asymptotic giant branch (AGB) stars throughout the Galactic Plane
that harbor thin disks and
mimic the SED of genuine Class III YSOs
\citep{Robitaille2008}.
Furthermore, carbon-rich red giant envelopes can also produce PAH emission at 8 $\mu$m
and therefore artificially increase the number of Class III candidates \citep[e.g.,][]{Strafella2015}.
We note that this classification scheme is based on what \citet{Greene1994} proposed,
 with the exception that we do not distinguish between ``flat spectrum'' and Class I types, but consider both as Class I.
This led to 13~968 YSOs that were classified as Class I and 28~317 as Class II.

However, the color criterion was developed by Gutermuth et al. by using deep {\it Spitzer}/IRAC imaging of nearby star-forming clouds, 
and studies have shown that significant false positives were
selected when this criterion was applied to GLIMPSE data 
of more distant star-forming regions \citep[e.g.,][]{Povich2009,Povich2013}.
Therefore, to further ensure a more robust sample of Class I and Class II candidates, 
we also identified the different classes using IRAC [3.6]-[4.5]/[5.8]-[8.0] color-color diagrams \cite[e.g.,][]{Allen2004}.
The 
IRAC color-color diagram with the YSO candidates and respective evolutionary classifications is displayed in Fig.~\ref{irac}.
We find that 10~694 ($\sim 77\%$) Class I and 18~209 ($\sim 64\%$) Class II
sources that are classified based on their spectral indices 
have the same evolutionary stage according to their position in the color-color diagram.
This YSO sample is considered in our further analysis.
 
\begin{figure}
\begin{minipage}{1\linewidth}
        \resizebox{\hsize}{!}{
        \includegraphics[angle=0]{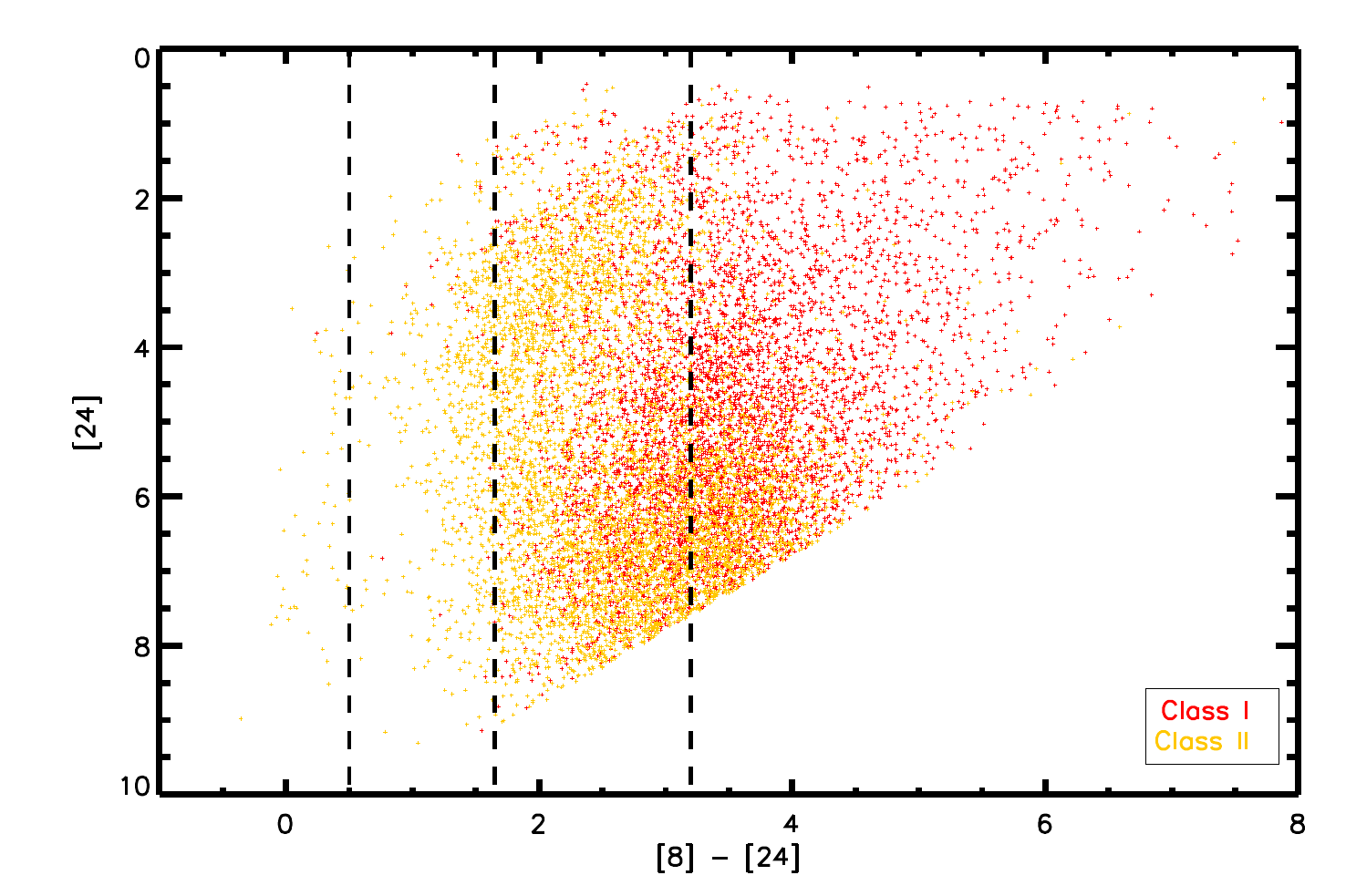}
        }       
\end{minipage}   
\caption{YSO candidates classified 
using the IRAC fluxes in a [24] versus [8]-[24] diagram. 
The vertical dashed lines indicate the expected regions for the different evolutionary stages, based on the [8]-[24] colors,
where Class I sources should have [8]-[24] $>$ 3.2, Class II sources 1.6 $\le$ [8]-[24] $\le$ 3.2, Class III sources
0.5 $\le$ [8]-[24] $\le$ 1.6,
 and main-sequence stars should have [8]-[24] $<$ 0.5.
}
\label{8vs24}
\end{figure}

Another concern arising from our YSO classification
 is the effect of extinction along the line of sight, which is
especially problematic in the Galactic Plane region. This effect may compromise some of the measured IRAC fluxes 
 and therefore affect our YSO classification, which is inferred from the spectral indices.
The reddening effect on sources that is due to extinction can potentially bias our YSO sample
toward the detection of false-positive YSOs 
and also move evolved YSO toward earlier evolutionary stages. 
In order to evaluate the effect caused by extinction, we
 performed an independent source classification using [8] and [24] for sources with a 24 $\mu$m counterpart in the MIPSGAL catalog,
since extinction is lower between 8 and 24~$\mu$m.
In Fig.~\ref{8vs24} we display the 12~997
 YSO candidates that have a 24 $\mu$m counterpart in a [8] vs [24] diagram.
By relating the SED spectral indices with the [8]-[24] color, we
reclassified our sources and compared the results with the $\alpha_{\rm IRAC}$ classification.

The results of this comparison, summarized in Table~\ref{tab:ysos},
show that the majority ($\sim$64\%) of the sources classified with the four IRAC bands 
maintain the same classification when reclassified using the [8]-[24] colors.
Although there is some uncertainty between the classes, 
it is reassuring 
to know that only a small fraction of YSOs 
($\sim0.4\%$)
are reclassified as pre-main-sequence stars in the 
[8]-[24] color diagram ([8]-[24] $< 0.5$).

However, 24~$\mu$m emission is usually diffuse or even saturated
in the inner regions of the bubbles 
because of 
the emission of small dust grains inside the ionized region of the bubble and the PDR.
Therefore, only $\sim$28\% (1~126 out of 4~044)
 of the YSO candidates located inside a bubble have a 24 $\mu$m counterpart,
while outside the bubbles we find $\sim$48\% (11~871 out of 24~859).
Nevertheless, more than 99$\%$ of the Class II and Class I YSO candidates 
with a 24~$\mu$m counterpart  
that were classified using $\alpha_{\rm IRAC}$
are still classified as a YSO after the [8]-[24] classification, which supports our assumption that 
the $\alpha_{\rm IRAC}$ classified sample
is not  significantly contaminated by stellar objects.

\begin{table}
\begin{minipage}{\linewidth}     
 \caption{
 Results of the comparison between the classification using $\alpha_{\rm IRAC}$ and [8]-[24] criteria for YSO candidates with 24~$\mu$m detection.
  }
\scriptsize
\begin{tabular}{c|c|c}
\hline\hline
\centering
\backslashbox{[8]-[24]$^{\rm (b)}$}{$\alpha_{\rm IRAC}+{\rm color}^{\rm (a)}$} & Class I (6157) & Class II  (6840)   \\
\hline
Class I & 3705 (60.2\%) & 1702 (24.9\%)  \\
\hline
Class II & 2423 (39.4\%) & 4543 (66.4\%)  \\
\hline
Class III & 28 (0.4\%) & 538 (7.9\%)  \\
\hline
Star & 1 (0.0\%) & 57 (0.8\%)  
\label{tab:ysos}
\end{tabular}
 \begin{list}{}{}
 \item[]{{\bf Notes:}  (a) YSO candidates with 24~$\mu$m detection classified by $\alpha_{\rm IRAC}$ criteria and color-color IRAC diagram \\
(b) YSO candidates classified using [8]-[24] colors.
}
\end{list}     
 \end{minipage}
\end{table}

\begin{figure*}[h!]
       \begin{minipage}{1\linewidth}
     \resizebox{\hsize}{!}{
   \includegraphics[angle=0]{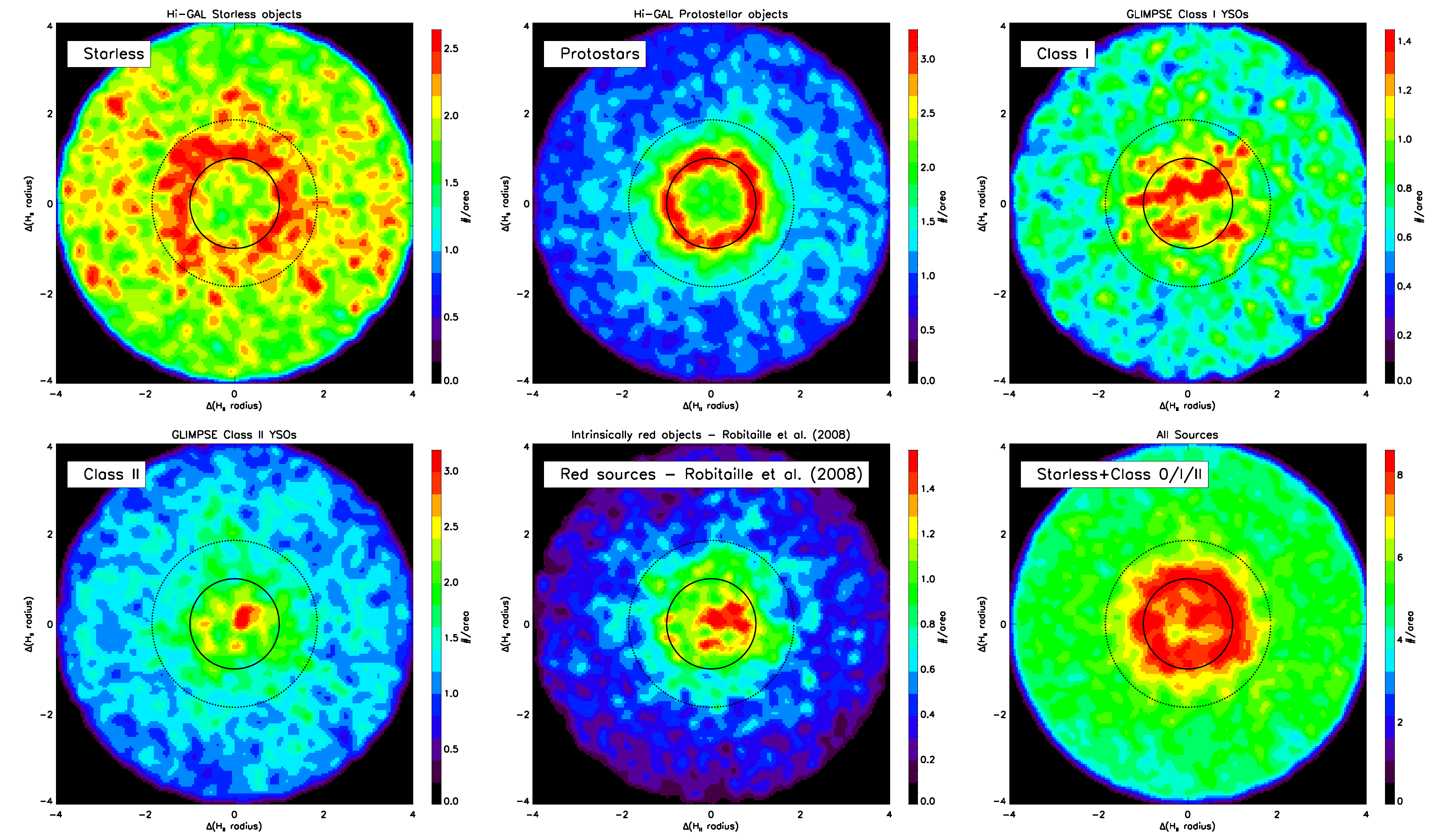}
   }
  \end{minipage}   
   \caption{Surface density maps
    for all star-forming objects - Hi-GAL clumps, IRAC YSOs, and intrinsically red sources from \citet{Robitaille2008} - associated with the bubble sample. The spatial scales of the maps are normalized by the bubble radius (solid black circle).
 The dashed black circle represents the average shell radius ($R_{\rm shell} = 1.85 \times R_{\rm eff}$).
The maps are smoothed by convolving with a Gaussian with a size of $FWHM = 0.25 R_{H_{\rm II}}$.
   }
              \label{HII_sources}
   \end{figure*}

\subsection{Dust temperature and column density distribution in the vicinity of \HII regions}

The four intensity maps from 160~$\mu$m to 500~$\mu$m were used to compute the column density and the dust temperature ($T_\mathrm{d}$) maps. 
First the three intensity maps at $\lambda<500$~$\mu$m were convolved with a Gaussian to degrade their spatial resolution: 
after this step, all the maps have an equivalent resolution of 36$\arcsec$ of the 500~$\mu$m (PLW) band; then the three maps were regridded onto the same spatial grid of the PLW image.
 After this step, the four maps can be stacked and $N(\mathrm{H}_2)$ and $T_\mathrm{d}$ can be derived by fitting 
 a modified blackbody pixel by pixel: it is assumed that the dust emission can be modeled as $I_\nu=\kappa_\nu\Sigma B_\nu(T_\mathrm{d}),$
 where $B_\nu(T_\mathrm{d})$ is the Planck function at temperature $T_\mathrm{d}$, 
 $\Sigma=\mu_{\mathrm{H}_2}m_{\mathrm H}N(\mathrm{H}_2)$ is the gas surface-density distribution 
 with $\mu_{\mathrm{H}_2}=2.8$ (molecular weight), $m_{\mathrm H}$ is the mass of the hydrogen atom, 
 and $N(\mathrm{H}_2)$ is the column density of molecular hydrogen. 
 The dust opacity per unit mass (assuming a gas-to-dust ratio of 100) is expressed as a power law: 
 $\kappa_\nu=0.1\times(300\mu{\mathrm m}/\lambda)^2$~cm$^2$/g. 
 The assumption on the molecular weight and on the functional form of $\kappa_\nu$ are typical of the Hi-GAL program (e.g., Elia et al. 2013)\nocite{Elia2013}.

The fitting procedure was executed with a C code that takes the four images as input.
The code creates a grid of models, as in \citet{Pezzuto2012}, according to the needs of the user:
since in our case the dust emissivity index $\beta$ is fixed to 2, the grid is built by varying only the temperature 
in the range $5\le T_\mathrm{d}(\mathrm{K})\le 50$ in steps of 0.01~K.
For each temperature, the grid gives the intensity at the four wavelengths corresponding to a column density of 1~g/cm$^2$,
and $I_\nu$ is linear in $N(\mathrm{H}_2)$ so that the column density for a given SED $f_i$
can be computed with a straightforward application of the least-squares technique. 
Then, for each pixel we have a set of pairs ($T_{\mathrm{d}_j},N(\mathrm{H}_2)$), with $j$ running over all the models in the grid;
 the pair with the lowest residuals is kept as best-fit model for that pixel. 
 Color corrections were not computed. The uncertainty associated to $I_\nu$ was assumed to be 20\% at all wavelengths. 
 More details on the production of the column density and temperature maps can be found elsewhere (Pezzuto et al., in prep; Schisano et al. in prep).

This method was produced to increase the computational speed 
needed in particular for the \herschel Hi-GAL fields that were
converted into larger mosaics.
Comparison with previous Hi-GAL column density maps following a different approach \citep[e.g.,][]{Elia2013,Schisano2014} 
reveals negligible inaccuracies between these two methods.
 
\section{Results} \label{sec:res}

     \begin{figure}
       \begin{minipage}{1\linewidth}
     \resizebox{\hsize}{!}{
   \includegraphics[angle=0]{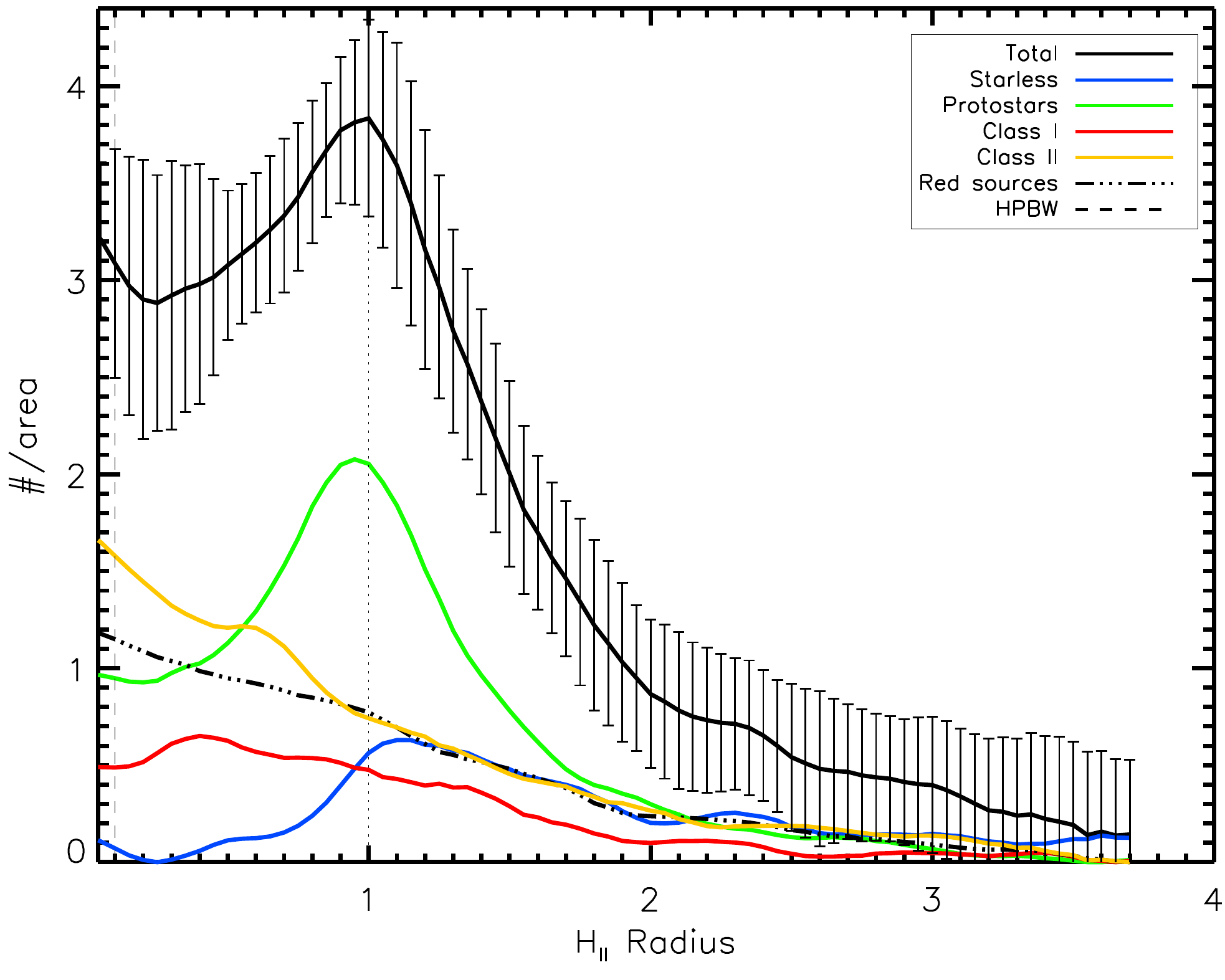}
   }
    \end{minipage}   
\caption{Mean radial source density profiles as a function of distance to the central position of the \HII regions,
obtained by azimuthally averaging from the surface density maps. 
The black solid line represents the cumulative count of the different source types (color-coded solid lines).
The black dash-dotted line represents the mean radial source density profile of the intrinsically red sources from \citet{Robitaille2008}.
 The distance is normalized by the radius of the \HII regions (dotted vertical line). The error bars show the ($\pm 1~\sigma$) dispersion of the radial profiles.
 The dashed vertical line represents the HPBW ($=0.125~R_{\rm eff}$) of the smoothing gaussian used in the surface density maps.  
   }
              \label{profiles}
   \end{figure}

\subsection{Distribution of star-forming sources related to ionized regions}\label{sec:sf_distribution}
The spatial information of all the associated star-forming objects
was compiled into surface density maps.
To take the different sizes of the bubbles  
into account, we normalized the  $l$ and $b$ offsets from the center of the bubbles
by the respective $R_{\rm eff}$ of the bubbles.
 A grid of 80 by 80 pixels with a pixel size of 0.05 $R_{\rm eff}$ was generated
 following the orientation of the Galactic coordinates, with
the x-axis aligned with the direction of the Galactic latitude $l$ and the y-axis with the Galactic longitude $b$. 
Then we measured the number counts for each star-forming object type at each pixel.
As a final step, the maps were smoothed by convolving them with a Gaussian with $FWHM = 0.25 R_{\rm eff}$.
In Fig.~\ref{HII_sources} we show the surface density maps for each star-forming object type.

We note that radial velocity measurements of YSOs and clumps that are spatially associated with bubbles have
been found to be consistent with ionized gas of \HII regions \citep[e.g.,][]{Martins2010,Hou2014,Deharveng2015,Liu2015,Liu2016}. 
We therefore assumed that star-forming objects that are found within the edges of bubbles
are physically associated with them.

The surface density maps provide a clear picture of
where
star-forming objects are typically located in the vicinity of bubbles.
These maps can be interpreted as probability function maps, 
revealing where star-forming objects at a given evolutionary stage are most probably located relative to a bubble.

Another visualization of the overdensity of star-forming objects is expressed as azimuthally averaged profiles of the surface density map,
as displayed in Fig.~\ref{profiles}. The surface density profile of all sources 
shows two significant peaks, one toward the center and the other toward the rim of the bubbles. 
The individual profiles 
 show 
that Class II YSOs dominate the inner regions of the bubbles,
 while younger starless and protostellar objects
 are mostly seen toward the edges.
In particular, the peak of each individual profile follows a gradient in age.

In order to analyze how significant the surface density of 
star-forming sources is ``inside'' the bubble and ``outside'' the bubbles, we first need 
to define the radius of the shells.
In previous studies the area up to 1.6 and 2 times the $R_{\rm eff}$ of the bubble
was considered as part of the bubble,
as the surface density of star-forming objects decreased to the level of the background \citep[e.g.,][]{Kendrew2012,Thompson2012}.
Here, however, we opted to take the different thicknesses of the bubbles into account.
Therefore, we defined the radius of the shell ($R_{\rm shell}$) of the bubble
as the sum of the angular thickness of the bubbles (provided 
from the MWP catalog) with $R_{\rm eff}$. 
On average, $R_{\rm shell}$ of our sample is $~\sim 1.85 \pm 0.21~\times~R_{\rm eff}$,
 which is approximately the same 
 as the constant inner radius assumed by previous authors.
Considering the surface density of all star-forming sources (excluding Class III) inside the shell ($< R_{\rm shell}$)
and outside ($R_{\rm shell} < r < 4 R_{\rm eff}$), we
estimate that the weighted mean ratio is 1.8$\pm$0.7, 
weighted by the total number of sources ($< 4 R_{\rm eff}$) per bubble.
This means that we typically find 80\% more star-forming objects 
per unit area 
toward the direction of bubbles than outside. 
Nevertheless, we note that some overlap between other bubbles
and the considered outer area ($R_{\rm shell} < r < 4R_{\rm eff}$)
 may contaminate our estimates of the surface density in these regions, 
especially for smaller bubbles, which are not considered here.
Although we expect this effect to be minor due to the large statistics,
this may slightly increase the number of sources located outside the bubbles.
Therefore we take the $\sim 80\%$ ratio as a lower estimate.

However, since the selection criteria used for the YSO candidates enforces
high-quality detection (magnitude errors $<$~0.2~mag) in all four IRAC bands,
a lack of detection of YSO candidates at the rim of 
the bubbles may be due to the very strong diffuse emission 
from the PDR at 8~$\mu$m.
To ensure that the overdensity of YSO sources in the inner part of the bubbles
is not biased for this reason, 
we dropped the requirement of having quality photometric measurements at [8.0] band 
and repeated our analysis 
by applying the Gutermuth et al. color criteria to identify YSOs. We also classified them into Class I and II using the spectral index slope and the color-color diagram, as in Sect~\ref{sec:ysos}.
To compensate for the poor quality of the photometric flux at 8.0 $\mu$m, we made
use of the 24~$\mu$m band measurement 
to evaluate the spectral index of the YSO candidates whenever available.

By applying these criteria, we obtained sample of 11~330 Class I and 27~879 Class II 
 YSO candidates,
which represents an increase of 6\% and 53\%, respectively, when compared with our previous criteria. 
In this sample we found that 6~125 ($\sim 54\%$) Class I and 7~448 ($\sim 27\%$) Class II candidates have a
measured MIPS [24] counterpart.
The resulting spatial distribution of sources, displayed in the online Fig.~\ref{HII_sources_3b},
shows almost exactly the same spatial YSO
distribution as we obtained with our previous criteria to construct Fig.~\ref{HII_sources}.
The increase in the number of YSO candidates further enhances the statistics of YSOs inside bubbles. 
This indicates that the overdensity of Class I and Class II YSOs in the inner part of the bubbles
are not an artefact that is due to the possibility of fewer detections caused by the PDR at the rim of the bubbles.

\onlfig{5}{
\begin{figure*}[h!]
       \begin{minipage}{1\linewidth}
     \resizebox{\hsize}{!}{
    \includegraphics[angle=0]{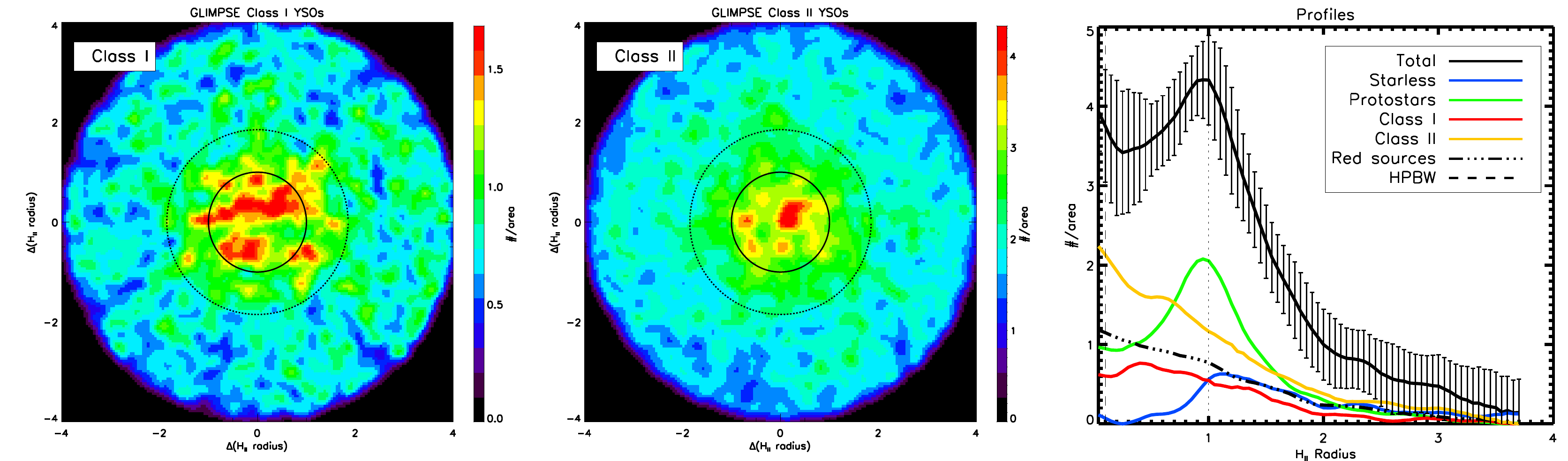}
       }
  \end{minipage}   
   \caption{Same as Figs.~\ref{HII_sources} and 4, preformed 
   for the YSO sample obtained by exempting quality detection at the IRAC [8.0] band and using the MIPS [24] photometric measurement  
   in the YSO classification when available. 
   }
              \label{HII_sources_3b}
   \end{figure*}
}

Last, to certify the reliability of the obtained result, we also analyzed in the same manner a more 
recent catalog of 
$\sim 133~980$ Class I/II YSOs candidates identified  
from the ALLWISE catalog \citep{Cutri2013} based on a vector machine selection \citep{Marton2016}.
We find that 46~761 of the sources lie in the same
area as is covered by the Hi-GAL survey of the inner Galactic plane. 
The WISE photometric data were combined with reliable 2MASS photometric measurements and Planck dust
opacity values. We found $\sim 24~517$ ($\sim 52.4\%$) sources
within the $< 4\times R_{\rm eff}$ criterion and 
 5~045 sources are located inside the bubble radii ($< R_{\rm eff}$).
In Fig.~\ref{wise}  we show the comparison of the surface density maps of the combined Class I and II YSO candidates
classified here in Sect.~\ref{sec:ysos} using IRAC fluxes and the Class I/II candidates taken from the WISE catalog. 
The YSO candidates of both selections peak at the center of the bubble with a surface unit of $\sim 4-5$ per bubble area (=$\pi~R_{\rm eff}^{2}$).
The main difference is mostly in the background sources with a lower value and in the flatter distribution of WISE YSO candidates.
That different observation sets and YSO classifications provide the same conclusion 
considerably strengthens this result.

\onlfig{6}{
\begin{figure*}[h!]
       \begin{minipage}{1\linewidth}
     \resizebox{\hsize}{!}{
   \includegraphics[angle=0]{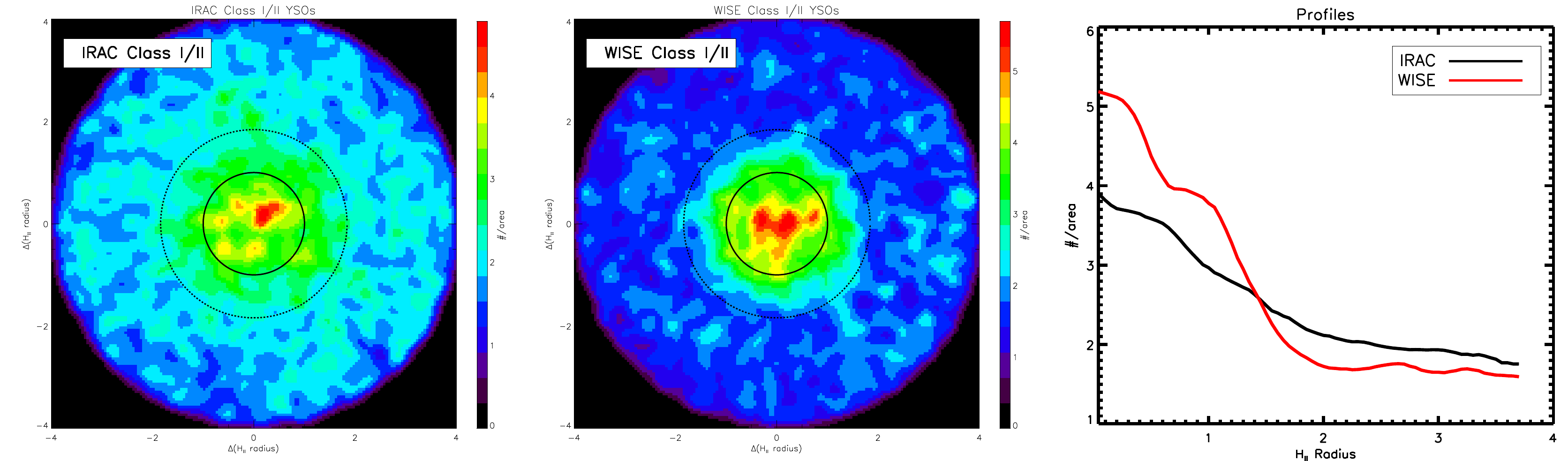}
   }
  \end{minipage}   
   \caption{Surface density maps for all Class I and II candidates classified using the four IRAC flux criteria in Sect.~\ref{sec:ysos} {\bf(left)}
    and candidates taken from the WISE catalog from \citet{Marton2016} {\bf(center)}.
   The spatial scales of the maps are normalized by the bubble radius (solid black circle). The dashed black circle represents the average shell radius ($R_{\rm shell} = 1.85 \times R_{\rm eff}$).
The mean radial source density profiles {\bf(right)} for the IRAC YSO candidates (solid black) and WISE YSO candidates (solid red) are shown as a function of distance to the central position of the \HII regions,
obtained by azimuthally averaging from the surface density maps. 
   }
              \label{wise}
   \end{figure*}
}

\subsection{Dynamic age estimates of \HII regions}\label{sec:ages}

The analytical solutions proposed by  \citet{Spitzer1978} and \citet{Dyson1980} are
commonly used to derive an age estimate by determining the required expansion time for an \HII region to reach its current size \citep[e.g.,][]{Zavagno2007}.
However, the predictions used in their method neglect the effect of the surrounding turbulent pressure,
which can influence the size of the \HII region, as demonstrated in \citet{Tremblin2012}.
Analytical solutions and numerical simulations performed in \citet{Tremblin2014}
demonstrated that the expansion of \HII regions is slowed down  by turbulent ram pressure ($P_{turb}$) 
of the environment
until it reaches quasi-static equilibrium with the pressure of the ionized gas ($P_{II}$).
These results allowed
Tremblin et al. to develop a reliable method to determine the dynamic ages of \HII regions.
 With the use of radio continuum flux measurements and by applying the Larson laws \citep[see][]{Larson1981} to infer $P_{II}$ and $P_{turb}$, respectively,
 dynamic age estimates were obtained by comparing the results with the isochrones provided by the grid of 1D models of expanding \HII regions.
The derived dynamic ages agree well with the photometric ages of the ionizing stars in well-known regions \citep[e.g., Rosette, RCW 36, RCW 79, and M16 $,\text{}$][]{Tremblin2014}. 
We found that 182 Galactic bubbles from our sample have a determined distance 
and associated radio flux measurements and are therefore eligible for determining a dynamic age using this method.
For this purpose, we made use of the 1.4 GHz radio continuum flux
measurements performed on the NVSS images, cataloged in 
 \citet{Condon1998}, for the northern part of the Galactic plane (120 found),
and the Parkes-MIT-NRAO (PMN) survey at 4.85 GHz \citep{Wright1994} for the southern part (62 found).
The distances of the Galactic bubbles were obtained from the  WISE catalog of Galactic \HII regions \citep{Anderson2014}.

Following \citet{Tremblin2014}, we first calculated 
the rms electron density 
$\langle n_{II} \rangle $ \citep[see][]{Martin-Hernandez2005}: 

\begin{equation}\label{eq_flux}
\langle n_{II} \rangle = \frac{4.092\times 10^{5}\mathrm{cm}^{-3}}{\sqrt{b(\nu,T_e)}}
\left(\frac{S_\nu}{\mathrm{Jy}}\right)^{0.5}\left(\frac{T_e}{10^4\mathrm{
    K}}\right)^{0.25}\left(\frac{D}{\mathrm{kpc}}\right)^{-0.5}\left(\frac{D_{\rm eff}}{''}\right)^{-1.5}
\end{equation}
\begin{equation}
b(\nu,T_e) = 1+0.3195\log\left(\frac{T_e}{10^4\mathrm{K}}\right)-0.2130\log\left(\frac{\nu}{\mathrm{GHz}}\right)
,\end{equation}
where $S_\nu$ is the radio continuum integrated flux at frequency
$\nu$ taken from the NVSS and PMN catalogs, $D_{\rm eff}$ is the angular diameter of the bubble, and
$D$ the
distance from the Sun.
 $T_e$ is the electron temperature in the
ionized plasma, and it was derived by using the
linear relation with galactocentric distance obtained by \citet{Balser2011}.
Then, the ionized gas pressure can be derived as follows:
\begin{equation}
 P_{II} = 2\langle n_{II} \rangle k_bT_e
.\end{equation}

  \begin{figure}
       \begin{minipage}{1\linewidth}
     \resizebox{\hsize}{!}{
   \includegraphics[angle=0]{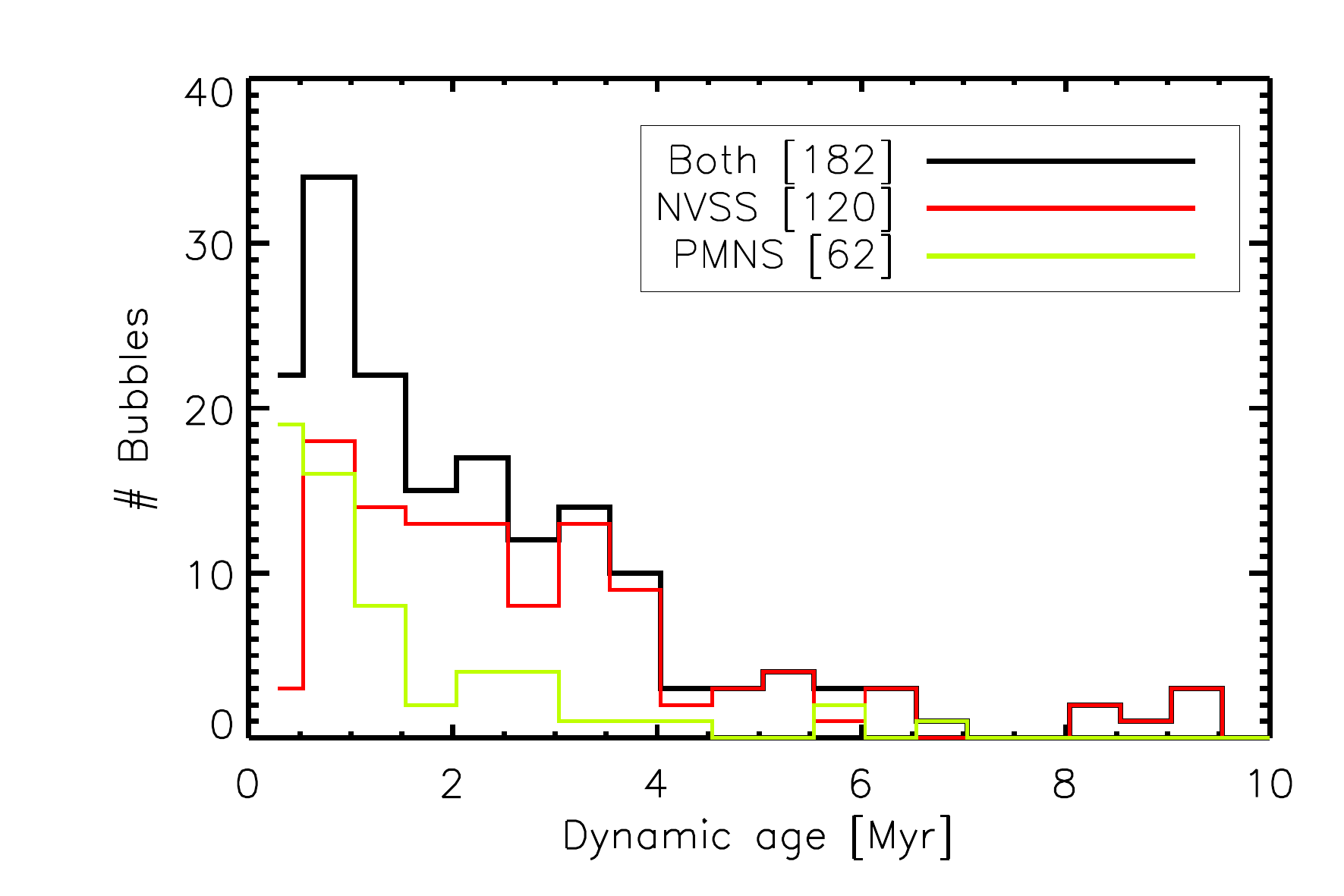}
   }
  \end{minipage}   
   \caption{Dynamic age histogram for a subsample of 182 bubbles estimated using the NVSS (in red) and PMNS (in green) surveys.
   }
              \label{hage}
   \end{figure}
 
  \begin{figure}
       \begin{minipage}{1\linewidth}
     \resizebox{\hsize}{!}{
   \includegraphics[angle=0]{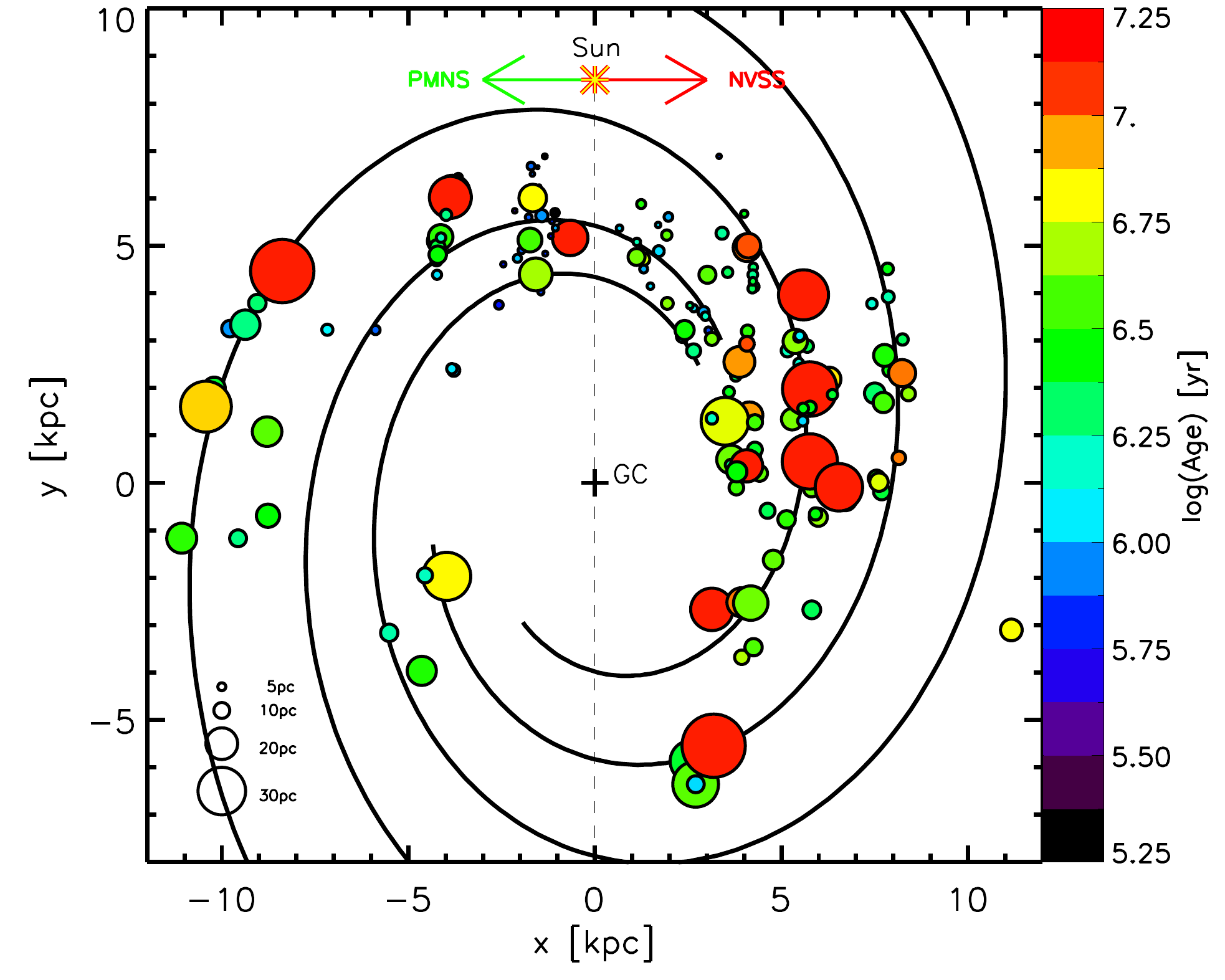}
   }
  \end{minipage}   
   \caption{Bubble distribution in Galactocentric coordinates with their respective diameter and age.
The black solid curves represent the position of the four Galactic spiral arms based on \citet{Russeil2003}. 
   }
              \label{gal_age}
   \end{figure}

Finally, we derive an estimate of the dynamic age for our subsample of 182 bubbles
by comparing it with the isochrones of the 1D simulations performed in \citet{Tremblin2014}. 
The distribution of the resulting ages is presented in Fig.~\ref{hage}.
We find a peak around 0.75 Myr in the histogram, 
and the large majority of the \HII regions have ages younger than 4 Myr ($\sim 80\%$).
Similarly, Tremblin et al. also found that most of
the derived dynamic ages for a sample of 119 regions 
that were detected using radio continuum measurements
obtained by the Green Bank Telescope at 9 GHz \citep{Anderson2011}
as part of the \HII Region Discovery Survey (HRDS)\footnote{http://www.cv.nrao.edu/hrds}
are younger than 4 Myr.
This distribution
 can be related with the typical lifetime of high-mass stars.
A main-sequence spectral type O5 star, for example, 
has an expected lifetime of $\sim 4$ Myr \citep{Allen1973}.

We note that the minimum angular size criteria of $R_{\rm eff} > 72\arcsec$ imposed for the MWP bubbles biases ages toward older bubbles.
As an extreme example, a bubble with an angular radius of $72\arcsec$ located at a far heliocentric distance 
of 15 kpc would correspond to a physical radius size of $\sim 5.2$~pc. 
For the Tremblin et al. models, this 
 corresponds to the expected size of a bubble created by an O5 star 
expanding into a 3~500~cm$^{-3}$ uniform medium during $\sim 2$ Myr.
This means that we might be excluding bubbles at
earlier stages of expansion (smaller radius) that are located at greater distance.
However, this criterion does not exclude the potential detection of older bubbles.
This age distribution may well reflect 
the typical lifetime of IR-bright \HII regions located in the Milky Way
to be around $< 4$~Myr,
for the majority of bubbles, or that IR-bright \HII regions are simply more visible during the earlier stages of expansion.

In Fig.~\ref{gal_age} we present a top view of the location of the 182 bubbles in the Galactic plane,
with their respective dynamic ages and sizes. 
In addition to finding more bubbles with an NVSS flux measurement in the northern Galactic plane (120)
than PMNS flux measurements in the southern Galactic plane (62), 
we also find bubbles at larger distances in the northern part.
The main reason for this is the lower detection limit of the NVSS survey 
($\sim 2.5$~mJy) compared with the PMNS survey ($\sim 20-40$~mJy).
Furthermore, the bubbles tend to be located in the Galactic spiral arms 
taken from \citet{Russeil2003}.

\subsection{IR emission from bubbles and shell density estimates}\label{sec:environment}

The multiwavelength images available throughout the Galactic plane, 
in particular the Hi-GAL, GLIMPSE and MIPSGAL surveys,
allow a thorough analysis of the local environment surrounding our sample of bubbles.

\begin{figure}
       \begin{minipage}{1\linewidth}
     \resizebox{\hsize}{!}{
   \includegraphics[angle=0]{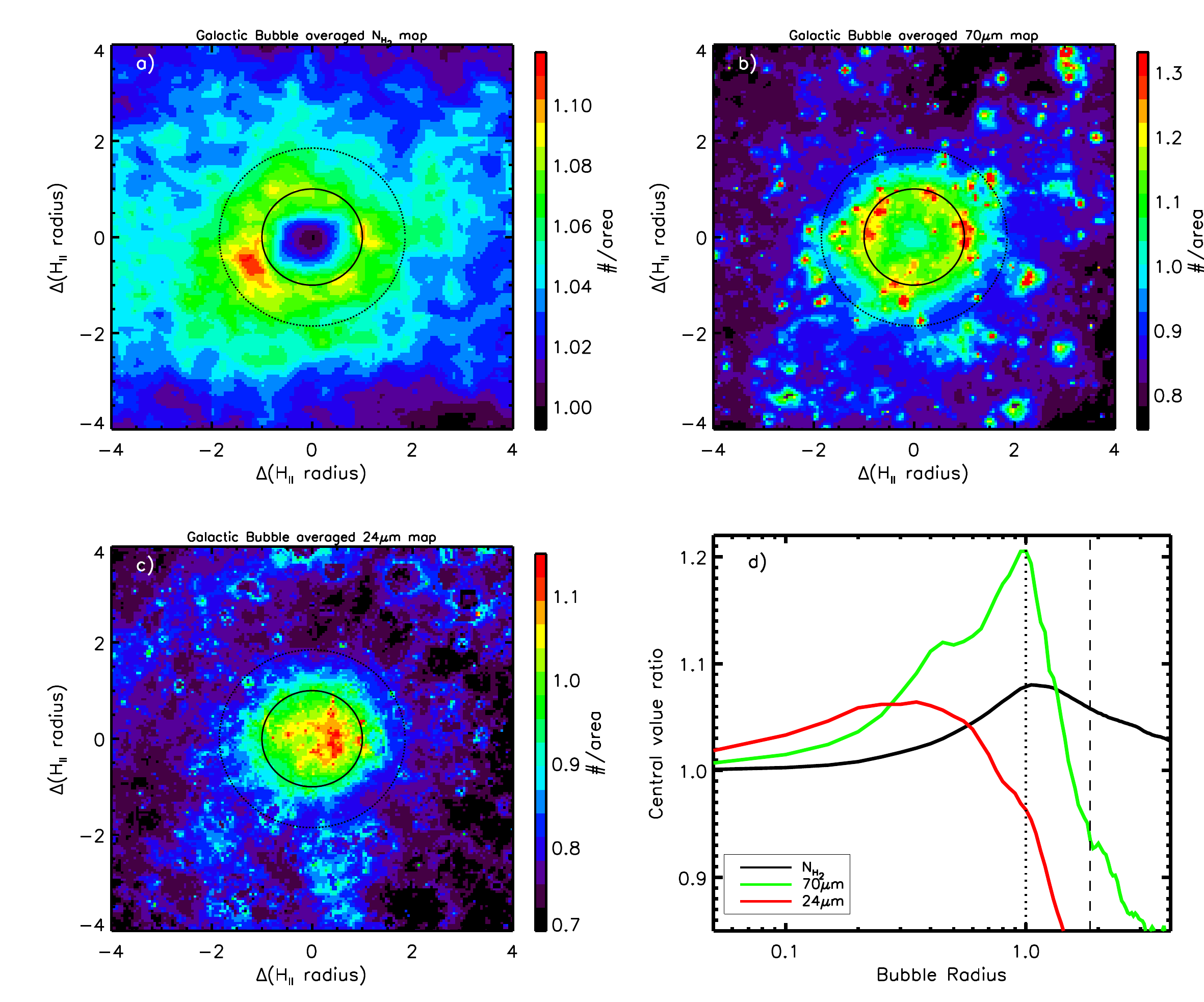}
   }
  \end{minipage}   
   \caption{Averaged column density and 70 $\mu$m and 24 $\mu$m maps of the Galactic bubble sample, {\bf a)}, {\bf b),} and {\bf c)}, respectively.
   The map units are normalized by the value in the central position of the bubbles.
   $R_{\rm eff}$ and $R_{\rm shell}$ are represented by the solid and dashed black circle, respectively.
   {\bf d)} Circularly averaged profiles of the averaged column density (black solid), 70 $\mu$m (green solid), and 24 $\mu$m (red solid) maps.
   The vertical dotted and dashed lines indicate $R_{\rm eff}$ and $R_{\rm shell}$, respectively.
   }
              \label{av_nh2_70}
   \end{figure}

Averaged column density and 70 $\mu$m and 24 $\mu$m maps were generated in order
to obtain an overview of the typical bubble environment.
For each bubble we considered the
squared area centered on the bubble of a length of $8 \times R_{\rm eff}$,
 normalized by its respective central value. 
The averaged maps were then generated by regridding all the maps to the same pixel size of $0.05 \times R_{\rm eff}$ 
and were averaged on a pixel-by-pixel basis
 for all the bubbles.
The resulting averaged maps and their respective azimuthally averaged profiles are displayed in Fig.~\ref{av_nh2_70}.
We note that the averaged bubbles are well resolved at the lowest resolution of the column density maps ($36.3\arcsec$)
and therefore no convolution was considered when performing the averaged maps. 

As expected, most of the 24 $\mu$m emission that traces the ionized gas is detected in the inner part of the bubbles,
while the 70 $\mu$m emission, which traces the ionized emission and the PDR, clearly peaks at the edges of the bubbles ($R_{\rm eff}$).
We also find that the column density is lower in the inner part of the bubbles and increases at the edge of the bubbles,
which is probably the effect of  material that is swept up by the expansion of the \HII region.

Furthermore, \herschel column density maps and bubble physical sizes allow us to estimate the mass of the bubble ($M_{\rm bub}$).
Using the 182 bubble sample from Sect.~\ref{sec:ages} for which we have kinematic distances,
we estimated the background for each bubble by taking the minimum value 
around a $8\times8~R_{\rm eff}$  box centered at each bubble.
A median filter of $35\arcsec$ size was applied before the background subtraction to avoid poorly SED fitted pixels of the column density maps.
After subtracting the local background (minimum value in the median filtered map), we calculated the mass for each bubble contained inside the 
shell of the bubble. 
We computed $M_{\rm bub}$ by summing all the pixels inside $R_{\rm shell}$ of the bubble, as follows: 

  \begin{figure}
       \begin{minipage}{1\linewidth}
     \resizebox{\hsize}{!}{
   \includegraphics[angle=0]{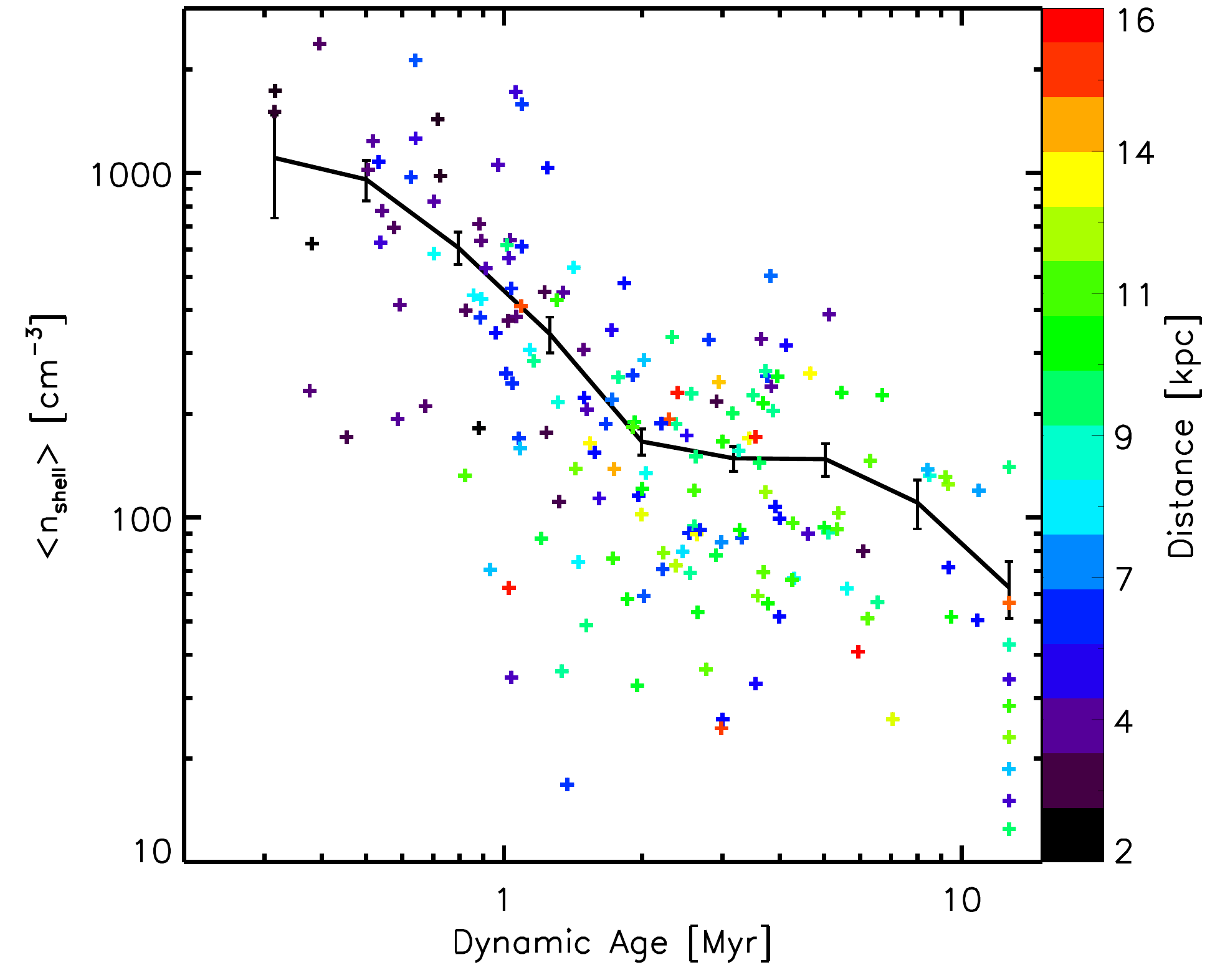}
   }
  \end{minipage}   
   \caption{Average number density inside the shell of the bubbles as a function of dynamic age of the bubbles. 
   The values are color-coded with the distance of their respective bubble.
   The solid line represents the average value, with the respective error bars representing the standard deviation of the mean. 
   The molecular cold gas located inside the shell ($R_{\rm eff} < r < R_{\rm shell}$) of the bubbles tends to be denser for younger bubbles.
     }
              \label{age_rho}
   \end{figure}

\begin{equation}
M_{\rm bub}~=~\sum~N_{\rm H_2}~\left(r < R_{\rm shell}\right)~\times~\left(\frac{A_{\rm pixel}}{\rm cm^2}\right)~\times~\mu_{H_2}~\times~m_{\rm H}~({\rm g})
,\end{equation}
where $A_{\rm pixel}$ is the pixel area in cm$^2$, $m_{H}$ is the atomic molecular of hydrogen, and
 $\mu_{H_2}$ is the mean molecular weight per $H_2$ molecule \citep[cf.][]{Kauffmann2008}.

Provided with mass estimates, we then calculated the mean volume density of the rim ($\rho_{\rm shell}$) by assuming that
the bubbles have a spherical geometry and that all the $M_{\rm bub}$ is confined
between $R_{\rm eff}$ and $R_{\rm shell}$, with no contribution from the inner part of the bubble.
In other words, we considered that $M_{\rm bub}$ was contained in the volume of the shell ($V_{\rm shell}$): 
\begin{equation}
V_{\rm shell} = \frac{4}{3}~\pi~\left({R_{\rm shell}}^3 - {R_{\rm eff}}^3\right)~(\rm cm^{-3})~.\end{equation}
Finally, the  $\rho_{\rm shell}$ was calculated by considering that the entire $M_{\rm bub}$ is contained inside the $V_{\rm shell}$.
In Fig.~\ref{age_rho}, we express our density estimates in the form of the average number density $<$$n_{\rm shell}$$>$  as a function of the dynamic age for 182 bubbles.
This plot shows a weak but significant trend between the mean density at the shell of the bubbles and their dynamic ages.
We find that typically, molecular gas concealed in the shells of the bubbles has a higher density at earlier stages of the bubble expansion,
which could therefore 
 suggest that bubbles are probably formed in high-density regions.
The decreasing trend would result from the increasing volume of the shell as the bubble expands 
and presumably converts part of its mass into stars.

\section{Discussion}\label{sec:disc}

\subsection{Star formation evolutionary gradient and timescales}\label{sect:evol}

The comprehensive statistical compilation of the spatial location of star-forming objects at different 
evolutionary stages, performed in Sect.~\ref{sec:sf_distribution}, 
reveals a clear evolutionary sequence. \citet{Thompson2012} indicated this possibility when they found that 
intrinsically red sources (more evolved sources) from \citet{Robitaille2008}
had a flat distribution inside the bubbles
(see Figs.~\ref{HII_sources} and \ref{profiles}),
while younger sources from the Red MSX Source (RMS) catalog \citep{Urquhart2008} were found to peak at the edges of the bubbles.

Interestingly, Class II YSOs in the inner part of the bubbles
are found for bubbles that have younger dynamic ages than 
the typical lifetime of 
low- and intermediate-mass
Class II objects 
$\sim~2~\pm~1$~Myr~\citep{Evans2009}.
This suggests that these YSOs have probably
undergone their formation process
before the expansion of the bubble,
possibly as part of the same star-forming complex that gave birth to the ionizing massive stars that are responsible for the expansion of the bubbles.

In cluster-forming environments, massive stars are expected to form after low-mass stars 
have completed their accretion phase \citep{Kumar2006}.
Moreover,
the well-studied case of the NGC~6357 complex, which is included in our sample of 1360 bubbles,
is a very good representative of this timescale issue.
Its relative proximity $\sim 1.7$ kpc \citep{Russeil2010}
allowed us to detect several Class II YSO ($\sim 220$) candidates
in deep IRAC observations around this complex, which are
found spatially distributed
around the $\sim 1$ Myr OB cluster that is ionizing the  \HII region \citep{Fang2012}.
Using less sensitive IRAC images, we find 64 Class II YSOs located in the inner part of the bubble ($< R_{\rm eff}$),
with a surface density $\sim 5.3$ times higher than the outer region in its proximity ($R_{\rm eff} < r < 4 R_{\rm eff}$).
The location of the Class II YSOs and the estimated age of the cluster
indeed suggest that these YSO were probably formed as part of the OB cluster that ionizes the region.

Moreover, a high-density environment of cold molecular gas is 
required for massive star-formation to occur,
which is a prerequisite for the formation of \HII regions.
This can be easily related with the
high-density shells of molecular gas found for bubbles 
at earlier stages of the bubble expansion (see Fig.~\ref{age_rho}).
Given that massive star formation is expected to be 
faster than low-mass star formation 
\citep{Mckee2007,Zinnecker2007},
it is probable that a few remaining sources were already undergoing formation 
at the moment the dense medium of the ``parent'' cloud started to become ionized. 
Furthermore, radiative magnetohydrodynamic simulations 
of expanding \HII regions show that denser clumps are more resistant 
to the expansion \citep{Geen2015}.
Although we did not find significant evidence for 
the existence of dense clumps inside the bubbles, 
it is possible that
photoevaporation by UV photons of the massive stars
partially dissipated their envelope because of their proximity \citep{Gritschneder2014}.

However, given the generally large distances of the bubbles and the sensitivity of the GLIMPSE data, most of the YSOs are probably intermediate- or high-mass objects and are therefore expected to have lifetimes shorter than 2 Myr. Therefore we also have to consider the possibility that the YSOs seen inside the bubble could have been formed through triggering by the bubble.
An interesting alternative is proposed in \citet{Gritschneder2014},
who presented a model of the dynamics of \HII regions and triggered stars.
The authors predict that triggered stars that formed inside the shell of the bubbles 
may migrate into the interior of \HII regions if the expansion occurs under
particular conditions, 
depending on the initial density of the environment, 
the formation time of the triggered star, and the flux of the ionizing star(s).
For cases such as Orion, the authors estimate that triggered stars may migrate back to the center of the \HII region within 0.6~Myr.
More importantly, their models predict that more evolved objects are seen 
closer to the ionizing star(s), following an age gradient.

Altogether, these arguments 
provide plausible 
scenarios
for the observed high
concentration of YSOs
in the interior of the bubbles
and the evolutionary trend in the surface density maps.

Nevertheless,
accurate measurements of proper motions
of YSO sources is crucial to understand this evolutionary sequence
toward the interior of bubbles for Class II sources.
When we can determine whether the YSOs are 
indeed migrating toward the center of \HII regions, we will be
able to provide a clear picture of how feedback 
interacts with the star-forming sources
and gain insight into the initial conditions of \HII regions.

For this reason,
we would like to point out that
high-precision infrared astronomic 
instruments, such as the 
Gemini Multi-conjugate Adaptive Optics System (GeMS) \citep{Rigaut2014},
can reach precisions on the order of $\sim 150~\mu$as \citep{Lu2014}
and could be used in the future to address this issue.
Furthermore, the instrument has previously been successfully 
used to study the RCW~41 \HII region 
\citep{Neichel2015}. 
In order to estimate the feasibility of
such a measurement,
we take
the expected velocity in the plane of the sky of $\sim 2$~km/s for a migrating
YSO close to the center of an \HII region from the model of Gritschneder
\& Burkert,
assuming an averaged inclination projection angle of 45$^{\circ}$.
For such a case we would expect for a \HII regions located at $\sim 2$~kpc distance
a proper motion velocity of $\sim 210$~$\mu$as/yr.
This is extremely encouraging 
for future proposals 
that would allow probing the internal dynamic motions
of star-forming objects inside bubbles. 

 \begin{figure}
       \begin{minipage}{1\linewidth}
     \resizebox{\hsize}{!}{
   \includegraphics[angle=0]{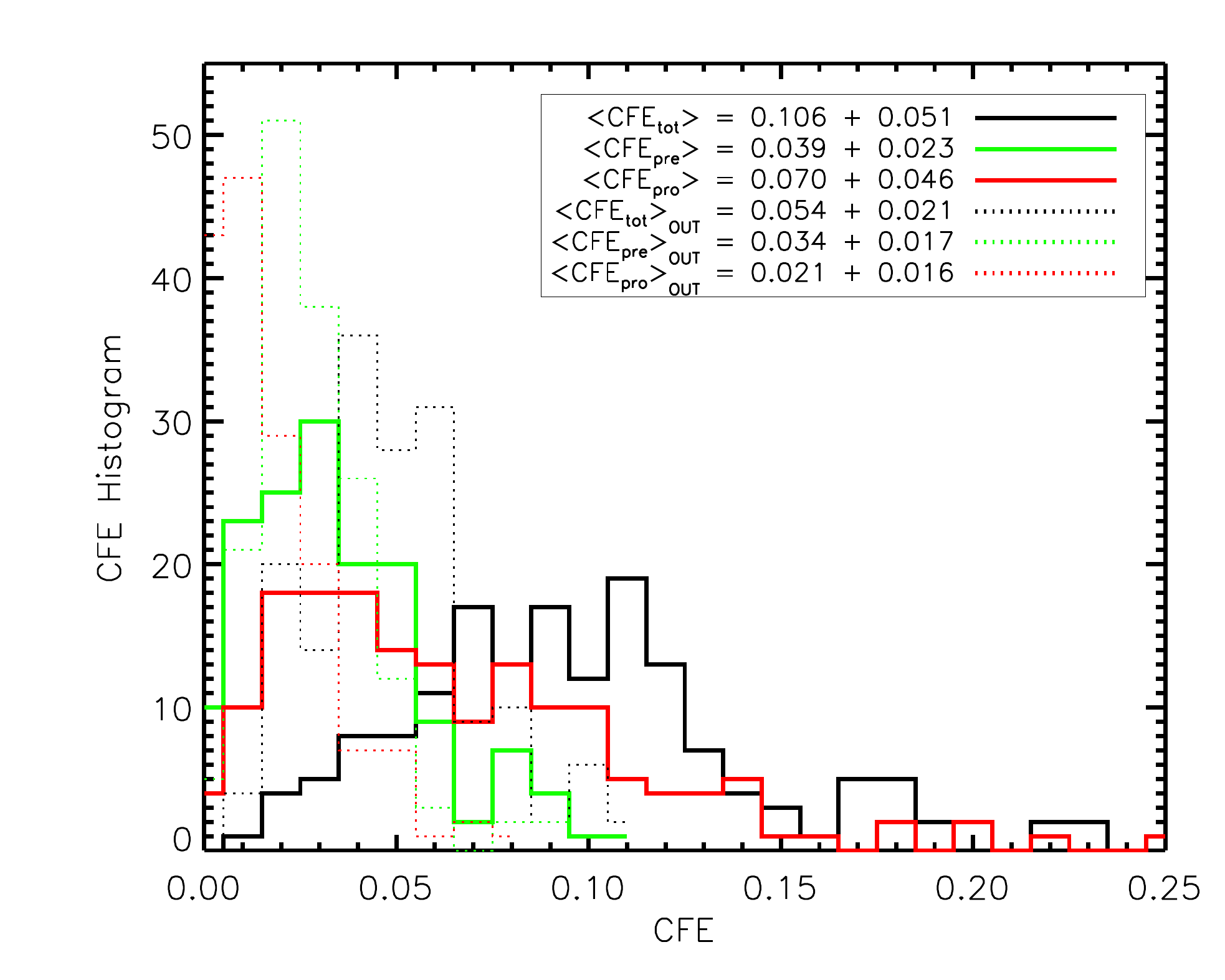}
   }
  \end{minipage}   
       \begin{minipage}{1\linewidth}
     \resizebox{\hsize}{!}{
   \includegraphics[angle=0]{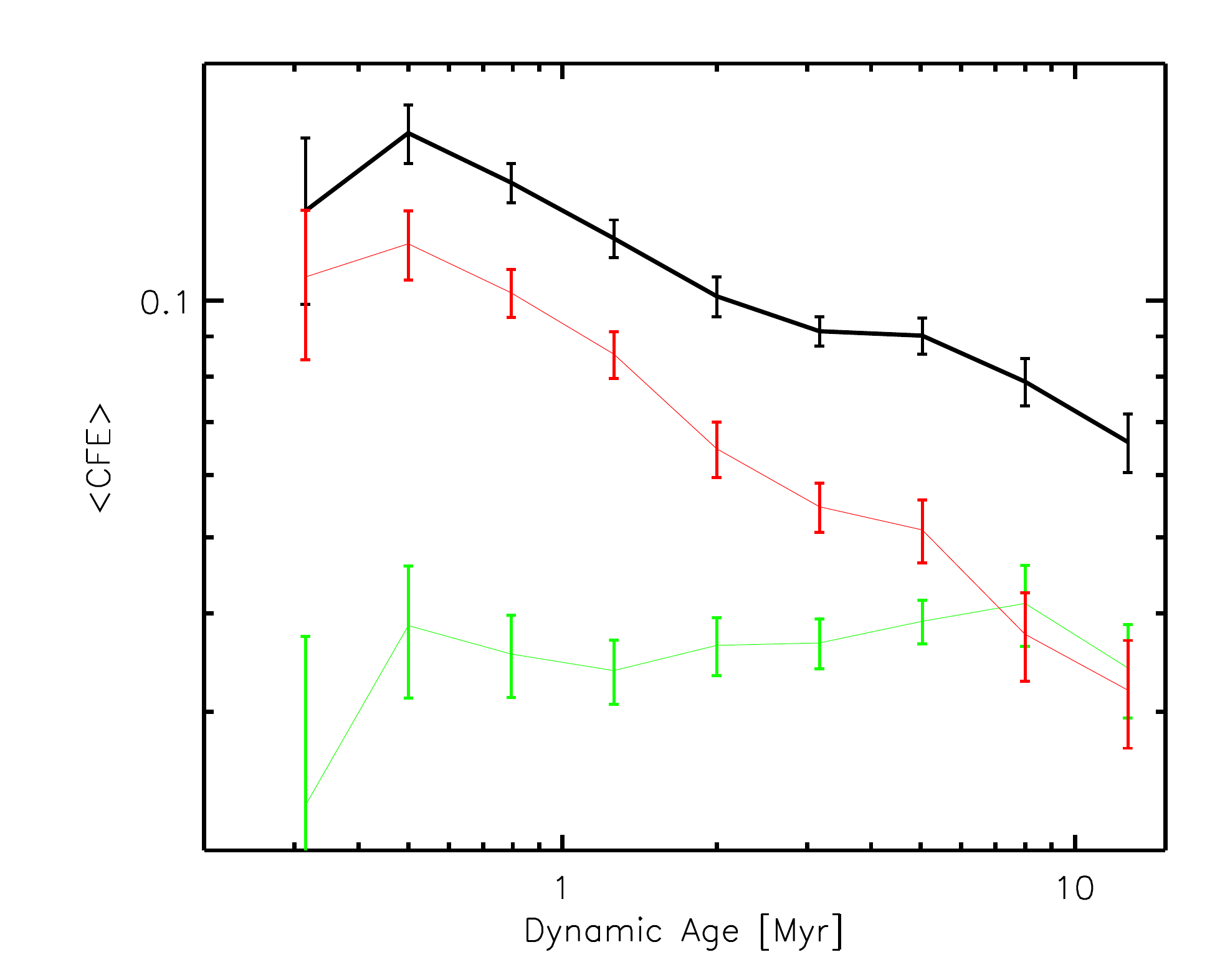}
   }
  \end{minipage}   
   \caption{{\bf (Top)} Histogram of the distribution of the total CFE (solid black), prestellar (solid green), and protostellar (solid red) sources
   inside the bubbles ($< R_{\rm shell}$) and outside ($R_{\rm shell} < r < 4 \times R_{\rm eff}$) (dotted).
{\bf (Bottom)} CFE as a function of the dynamic age of the bubbles.
   }
              \label{CFE}
   \end{figure}

\subsection{Clump formation efficiency of bubbles} 

A crucial aspect of the 
overall impact of \HII regions on the star formation process 
is to understand how efficient the conversion 
of the neutral matter collected in their layers
is compared with typical 
regions that are unaffected by feedback processes.
Numerical simulations have found that 
the net amount of star formation
might be lower for \HII regions,
since large portions of neutral 
material are expected to be dispersed 
by the ionizing star(s) \citep{Dale2012,Walch2012}.
The overdensity of star-forming clumps
near bubbles has been significantly well established 
\citep[e.g.,][]{Deharveng2010,Hou2014,Kendrew2016}.
These observation have given support to the C\&C theory 
as the main physical mechanism responsible for the 
formation of clumps, where 
swept-up material is compressed by the expansion,
fragments, and forms clumps through gravitational instabilities \citep{Elmegreen1977,Whitworth1994}.
However, because we find more cold dust in these locations,
it is essential to estimate how star formation processes inside the shells differ from 
processes in other regions that are unaffected by feedback effects.

An important quantity to address in this context is therefore
to infer how efficiently the cold molecular gas in
the shells is converted into star-forming clumps. We do this by estimating the 
clump formation efficiency (CFE) of the bubbles \citep{Eden2012,Eden2013}.
The CFE can be determined 
by calculating the ratio between 
the masses of the bubbles ($M_{\rm bub}$), calculated in Sect.~\ref{sec:environment},
 and their respective associated Hi-GAL sources (prestellar and protostellar clumps).

To be more consistent,
 we decided to estimate
 the envelope mass of the Hi-GAL sources in the same manner as $M_{\rm bub}$ was calculated
 (i.e., directly form the Hi-GAL column-density maps) to provide a more reliable estimate of the CFE.
 In this manner, mass uncertainties that are due to incorrect distance estimates, unreliable SED fits, or inaccurate background subtraction can be 
 significantly attenuated.
  Envelope masses of Hi-GAL prestellar ($M_{\rm pre}$) and protostellar ($M_{\rm pro}$) clumps
  located inside bubbles ($< R_{\rm shell} $)
 were calculated by integrating the column density map 
 over their circularized FWHM  sizes\footnote{
 We convolved the deconvolved clump sizes (${FWHM_{\rm dec}}$) provided in the Hi-GAL physical catalog
into the resolution of the column density map 
 ($\sqrt{{FWHM_{\rm dec}}^2 + 36.3\arcsec^2}$).
 }.
This provided us with estimates of 
the CFE of prestellar clumps (CFE$_{\rm pre}$ = $M_{\rm pre}$/$M_{\rm bub}$) and protostellar clumps (CFE$_{\rm pro}$ = $M_{\rm pro}$/$M_{\rm bub}$)
 for the sample of 182 bubbles we used to estimate dynamic ages in Sect.~\ref{sec:ages}.

We obtained considerably high values of CFEs for both prestellar, $<$CFE$_{\rm pre}$$>$ = $0.039 \pm 0.023$
 and protostellar $<$CFE$_{\rm pro}$$>$ = $0.070 \pm 0.046$ clumps.
The distribution of CFEs is shown in the histogram of Fig.~\ref{CFE}. 
For the total sum of prestellar and protostellar clumps we obtain $<$CFE$_{\rm tot}$$>$ = $0.106 \pm 0.051$, 
which means that typically $\sim 10\%$ of the molecular gas around the bubbles are concentrated in the form of prestellar or protostellar clumps.
For comparison, \citet{Elia2013} estimated a CFE $\sim 0.02 - 0.07$ for both protostellar and starless (including gravitationally bound and unbound) Hi-GAL clumps  
in four regions located in the outer part of the Galactic plane ($217^{\circ} < l < 224^{\circ}$). 
However, since gravitationally unbound clumps are not included in 
our analysis,
the difference between the CFE that we obtained with the CFE of these four regions
is probably higher.
Furthermore, Elia et al. estimated the mass of the considered regions using NANTEN CO$(1-0)$ observations, which
makes this comparison less reliable.
  
In order to make a more consistent comparison,
we also evaluated the CFE outside the bubbles (CFE$_{\rm out}$) between $R_{\rm shell} < r < 4 \times R_{\rm eff}$.
This was made by evaluating the ratio between the
total amount of integrated column density inside clumps (prestellar and protostellar) 
located in the outer region
and the total integrated area in the outer regions.
The dotted color lines in Fig.~\ref{CFE} show that the values of CFE$_{\rm out}$ are systematically lower 
than the CFE calculated for clumps inside the bubble shells, by an average factor of $\sim 1.97 \pm 1.21$. 
Moreover, comparing these values with 
well-known active star-forming regions, 
we find that the CFE inside the bubbles is also twice the value
obtained for RCW106 \citep{Nguyen2015} and W43 \citep{Nguyen2011}, and
approximately 10 times higher than Cygnus X \citep{Motte2007}.

Provided with estimates of the bubble dynamic ages (Sect.~\ref{sec:ages}), 
we determine possible variations of the CFE with the evolution of bubbles (bottom Fig.~\ref{CFE}).
Surprisingly, we find that CFE$_{\rm pro}$
tends to decrease with the age of the bubble, while CFE$_{\rm pre}$ seems to remain nearly constant.
A more careful interpretation, however, is required to fully understand the possible reasons for this trend. 
For instance, we need to consider that at large distances we might  
on one hand 
obtain fewer protostellar clumps because of the 
sensitivity limit of the \herschel PACS 70 $\mu$m band,
and on the other hand, we might obtain more source confusion with overlapping prestellar clumps 
 (see Elia et al. 2017).\nocite{Elia2017}
Based on Spearman's rank correlation, we find a moderate 
 correlation between the kinematic distances and the estimated dynamic ages (see also Fig.~\ref{age_rho})
 with a coefficient of $r_{\rm s} \sim 0.46$ and a p-value of $\sim 10^{-11}$.
Therefore we should consider the possibility that the decreasing trend of CFE$_{\rm pro}$ with age 
could be the result of fewer detections at farther distances that is due to sensitivity.
However, we find no evidence for a correlation between the kinematic distances and CFE$_{\rm pro}$ when applying the same test
($r_{\rm s} \sim 0.03$; p-value $\sim 0.66$), while the correlation between the dynamic ages of the bubble and CFE$_{\rm pro}$ seen in Fig.~\ref{CFE}
is statistically significant ($r_{\rm s} \sim -0.33$; p-value $\sim 10^{-5}$).

Moreover, the different trends between prestellar and protostellar 
 CFEs could be an indirect indication of changes in their formation rate as bubbles expand. 
A possible interpretation of these different trends 
might be a faster conversion between the prestellar to protostellar phase
for younger bubbles, which conversion gradually decreases as the expansion decelerates and 
the ionization radiation has a weaker impact.
Nevertheless, a detailed individual analysis of these bubbles is required 
to prove changes in the formation rate between prestellar and protostellar clumps. 
If the formation rates in \HII regions are similar to those of typical star-forming molecular clouds,
we expect that the net amount of star formation is higher.
We therefore suggest that
\HII regions may
have a positive impact 
on the star formation process throughout our Galaxy.

We estimate that $\sim 23\%$ of the Hi-GAL star-forming clumps 
($\sim 15\%$ and $\sim 41\%$ for prestellar and protostellar clumps, respectively) 
located in the inner part of the Galactic plane
are found projected toward a bubble ($< R_{\rm shell}$).
We find this value to be consistent with the $\sim 25\%$ ATLASGAL clumps
found by \citet{Kendrew2016}
from studying MWP bubbles.
It is particularly interesting to note the higher fraction of protostellar clumps.
A possible interpretation for the higher number of protostellar objects might be a higher formation rate that
converts prestellar clumps into protostellar clumps faster, possibly as a result of a feedback effect (through the C\&C or RDI mechanism).
Overall, these numbers suggest that the formation of many stars in the Galaxy
may have been triggered by (H{\sc{ii}}) bubbles.

\subsection{High-mass star formation in bubbles} 

Statistical studies of massive MYSOs
and their relation to bubbles
have been recently made.
In particular, 
\citet{Thompson2012} and \cite{Kendrew2012}
presented a statistical study
showing that a significant overdensity of MYSOs 
was identified in the Red MSX Source (RMS)
survey \citep{Urquhart2008}, 
near the rims of bubbles.
When we take the typical lifetime of a few $10^5$ yr \citep{Mottram2011}
 into account, these object trace very recent star formation.
In addition to proposing that
MYSOs were triggered by the expansion of (H{\sc{ii}}) bubbles,
both studies found evidence that bubbles associated with RMS MYSOs
tend to have smaller angular sizes,
which suggests that MYSOs tend to form around younger bubbles.

\begin{figure}[h!]
       \begin{minipage}{1\linewidth}
     \resizebox{\hsize}{!}{
  \includegraphics[angle=0]{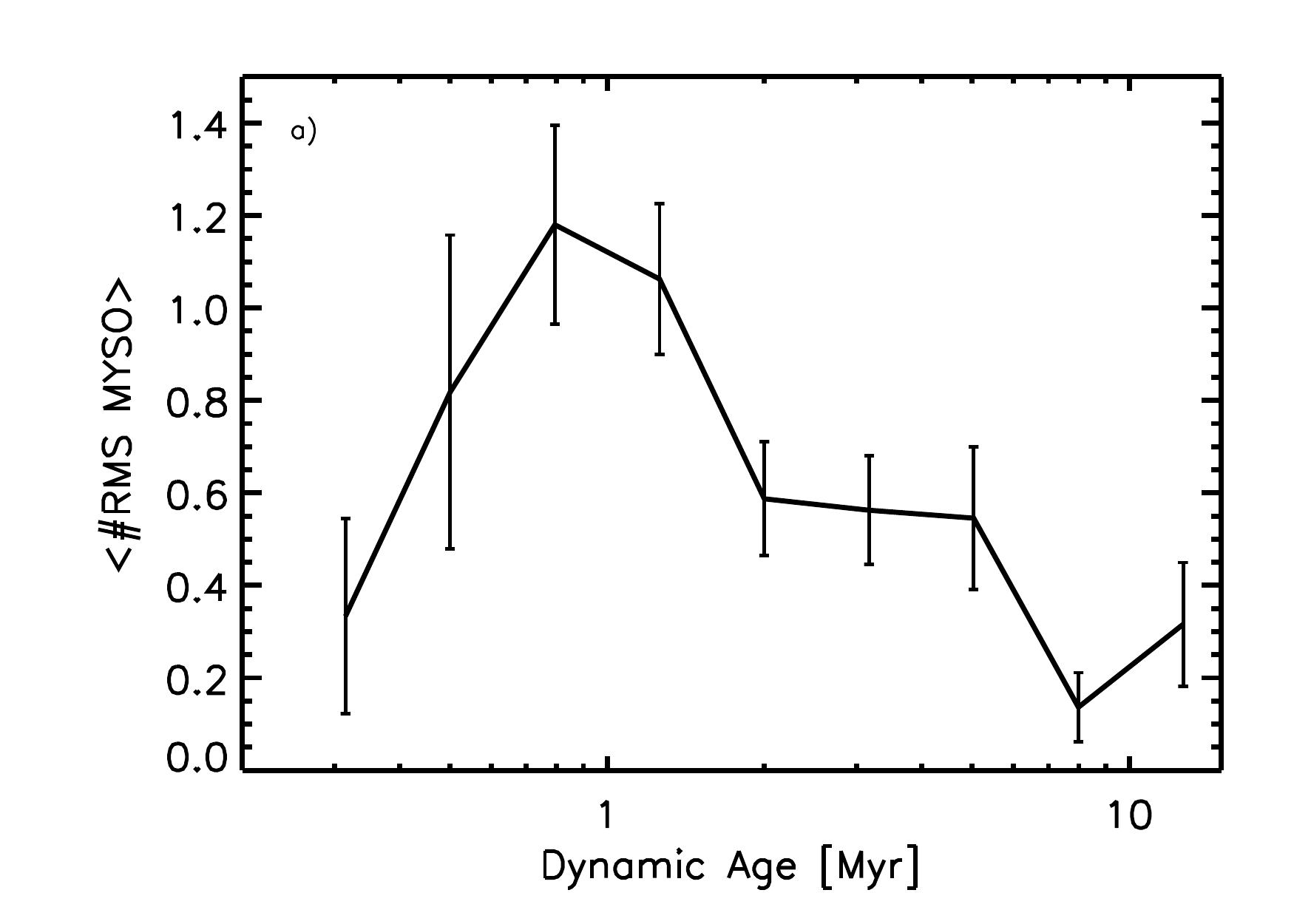}
   }
  \end{minipage}   
       \begin{minipage}{1\linewidth}
     \resizebox{\hsize}{!}{
  \includegraphics[angle=0]{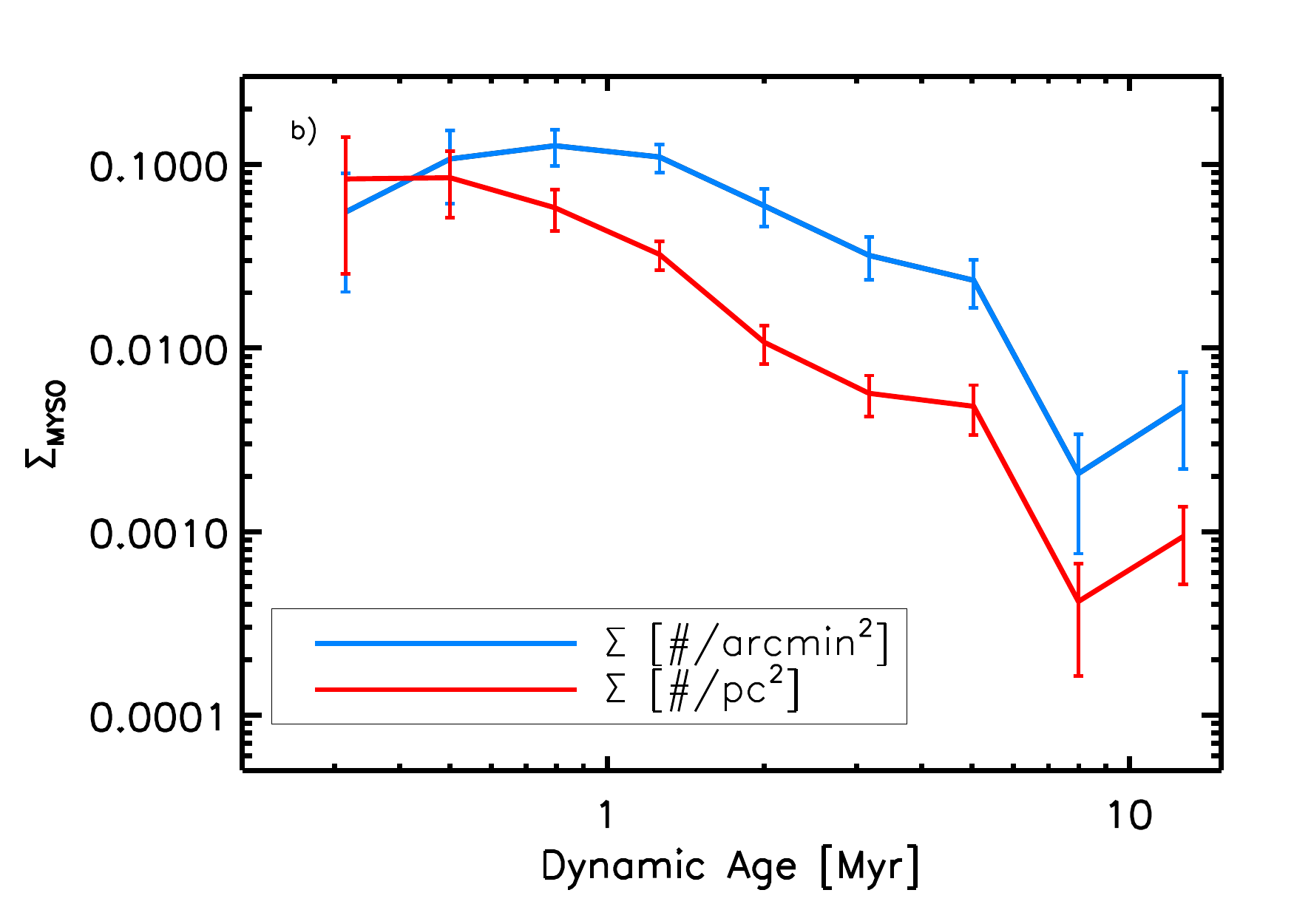}
   }
  \end{minipage}   
       \begin{minipage}{1\linewidth}
     \resizebox{\hsize}{!}{
   \includegraphics[angle=0]{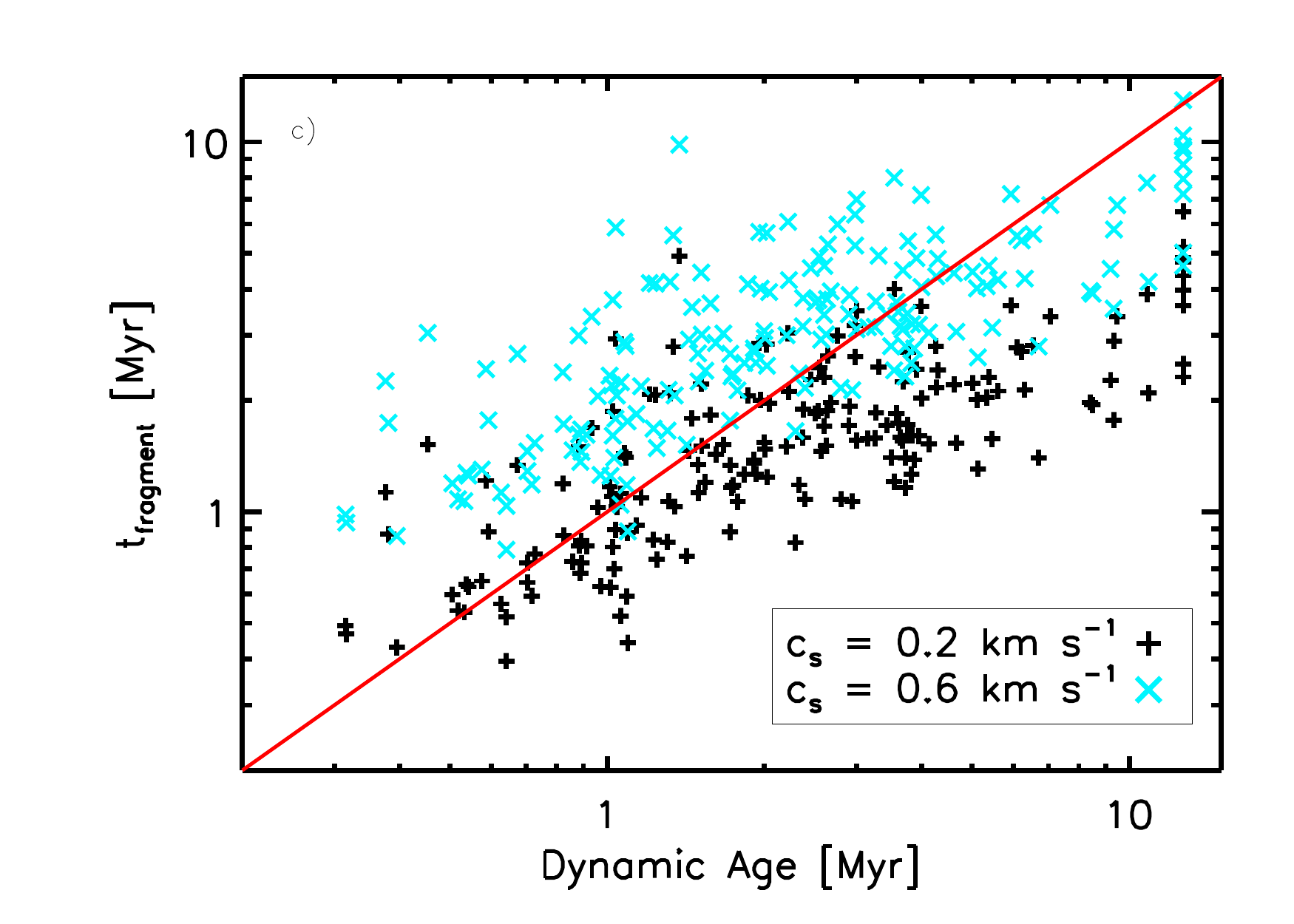}
   }
  \end{minipage}   
   \caption{{\bf (a)} Average number of RMS sources as a function of dynamic age.  
   {\bf (b)} Average apparent (in blue) and physical (in red) surface densities of RMS sources as a function of dynamic age.
The error bars show the standard deviation of the mean ($\pm 1~\sigma/\sqrt(N)$).
{\bf (c)} Fragmentation time as a function of dynamic time \citep{Whitworth1994},
assuming $c_{\rm s} = 0.2$ (in black) and 0.6 km s$^{-1}$ (in blue). The solid red line represents the one-to-one relation.
   }
              \label{MYSOs}
   \end{figure}

In order to shed some light into this subject,
we studied the potential association of RMS MYSOs 
with the 182 MWP bubbles for which we 
have derived the dynamic age.
We find that nearly 40\% (71) of these bubbles are associated with at least one MYSO.
We then analyzed the average number 
of MYSOs associated with bubbles ($< R_{\rm shell}$)
and searched for possible trends 
with the dynamic ages of the bubbles (see Fig.~\ref{MYSOs}a).
Interestingly, we find a higher fraction of younger bubbles ($< 2$~Myr) that are 
associated with at least one RMS MYSOs and also with higher numbers.
Around 50\% of the bubbles younger than 2~Myr are associated with RMS MYSOs (43 out of 86)
with an average number of $2.0\pm 1.1$ MYSOs, while only 29\% of bubbles older than 2~Myr 
have associated MYSO (28 out of 96) with an average of $1.5\pm 1.2$ MYSOs.

This is also evident from the average surface density of RMS MYSO,
using both apparent and physical sizes,
as a function of dynamic age in Fig.~\ref{MYSOs}b.
This result suggests that
the timescale for the 
formation of 
MYSOs in bubbles 
is probably fairly short
and very effective 
at the early stages of the bubble expansion.
We note that the RMS source catalog, which is
composed of sources with L$~> 10^4$~L$_{\odot}$
, is expected to be complete up to 
distances $\sim 15$~kpc \citep{Thompson2012},
which includes all the distances attributed to our subsample of bubbles.

Given the short time of expansion of the bubbles, 
the question arises as to whether 
the MYSOs were already present (or undergoing formation) before the formation of 
the (H{\sc{ii}}) bubbles or if they had the time to form through gravitational instability of the shell (C\&C process).
To search for a potential answer, we
estimated the fragmentation time for the shell of these bubbles 
using the analytical framework of \citet{Whitworth1994},
in order to relate the results with the dynamic ages of the bubbles.
Following Whitworth et al., the time at which fragmentation starts is determined by the following relation:
\begin{equation}
t_{\rm frag} \sim 1.56~{\rm Myr}~\left({\frac{c_{\rm s}}{0.2~{\rm cm~s^{-1}}}}\right)^{\frac{7}{11}} \left({\frac{N_{\rm Lyc}}{10^{49}~{\rm s}^{-1}}}\right)^{\frac{-1}{11}} \left({\frac{n_{0}}{100~{\rm cm}^{-3}}}\right)^{\frac{-5}{3}}
,\end{equation}
where $c_{\rm s}$ is the uniform isothermal sound speed,
$N_{\rm Lyc}$ is the Lyman-continuum photon emission
and $n_{0}$ is the initial mean number density of the medium before the expansion of the \HII region.
To obtain an estimate of $n_{0}$
, we assumed that the entire estimated mass contained in the shell
of bubbles was previously uniformly distributed inside the
total volume of the bubbles ($\le R_{\rm eff}$).
This resulted in an average $<$$n_{0}$$>~\approx 2~500$ cm$^{-3}$ ranging from 100 to $1.8\times10^4$ cm$^{-3}$
for our sample of 182.
$N_{\rm Lyc}$ was derived
from the NVSS and PMNS radio continuum fluxes (see Sect. \ref{sec:ages}).
An isothermal sound speed of the medium ranging from 0.2 to 0.6 km~s$^{-1}$ was also assumed.

In Fig.~\ref{MYSOs}c we display the estimated shell fragmentation times against
the dynamic ages of the bubbles. 
Assuming $c_{\rm s} = 0.2$~km~s$^{-1}$ 
, we find that 132 bubbles ($\sim 73\%$) have fragmentation times shorter than 
their predicted dynamic age.
However, we find that the number of bubbles
would decrease to 53 ($\sim 30\%$) when 
$c_{\rm s} = 0.6$~km~s$^{-1}$ is 
assumed.

According to this analysis, a significant fraction of these bubbles 
might not have had time to fragment,
therefore we should consider whether the associated MYSOs 
existed before the bubble began to expand or possibly were already undergoing formation.
This question has been discussed before.
As an example, 
\citet{Dewargan2013} performed a detailed study of the N14 bubble (from the Churchwell et al. catalog)
and estimated a shorter dynamic age than the
predicted fragmentation time.
Numerical simulations 
also show that preexisting dense structures in a fractal medium can 
further enhance their density through expansion of the bubble,
which leads them to collapse faster than they originally would 
\citep{Walch2012,Walch2015}.
Furthermore, 
star formation is predicted to occur around $\sim 0.5$ Myr
after the expansion of the \HII region \citep{Walch2013},
which agrees with 
our observations and dynamic age estimates.

These points 
indicate
that the initial environment 
surrounding an \HII region 
is probably not well described by
a neutral medium with uniform density.
As discussed in Sect.~\ref{sect:evol},
the presence of more evolved YSOs inside
bubbles suggests that the medium was actively forming stars at the moment
the massive star(s) started to ionize the medium,
indicating that density fluctuation were probably present in the medium.
Furthermore,
the high-density environment of the bubbles at early stages of their expansion
and their high CFE are favorable physical conditions for a more efficient formation of MYSOs
during the earlier stages of bubble expansion ($< 2$ Myr).

We note that a Galactic plane bubble catalog has recently been produced within the VIALACTEA consortium.
Bubble identification was obtained based on localized active contours that allow tracing the heated dust of bubble structures 
seen in emission in the Herschel 70~$\mu$m images 
and to define their morphology in the form of contours (Carey et al. in prep)\footnote{The active contour algorithm has been developed in the University of Leeds.}.
The additional information of the morphology of the bubbles and their inhomogeneities will allow us to further improve our understanding 
of the impact of ionizing stars on the interstellar medium.

\section{Conclusions}\label{sec:conc}
 
We carried out an extensive statistical study of star-forming 
objects in the vicinity of Galactic bubbles and analyzed their local environment.
The range of evolutionary stages of the star-forming objects, from prestellar clumps to 
Class II YSOs, provided a clear and distinct picture of the spatial distribution of star-forming sources 
in the surroundings of 1~360 (H{\sc{ii}}) bubbles.
Prestellar and protostellar clumps were taken from the Hi-GAL catalog,
while YSO candidates were classified using {\it Spitzer} IRAC fluxes.
This led to a total of $\sim 70~000$ star-forming objects that were compiled into
surface density maps. These revealed a clear evolutionary trend, 
where more evolved sources (Class II YSO candidates) are 
mostly spatially found concentrated near the center of the bubble,
while younger sources (HI-GAL prestellar and protostellar clumps)
are found at the rim.
We find $~80\%$ more sources per unit area
in the direction of the bubbles ($< R_{\rm shell}$) 
than outside the bubbles ($R_{\rm shell} < r < 4~R_{\rm eff}$). 
We evaluated and excluded the possibility of potential biases due to the YSO classification method
by using different classification methods. 
Furthermore, YSO candidates from a recent
WISE catalog \citep{Marton2016} also show a clear overdensity of
YSO sources toward the inner part of the bubbles, strengthening this result.

We were able to determine the dynamic age for a subsample of 182 bubbles
for which we found available kinematic distances and radio continuum flux measurements.
Most of the bubbles ($\sim80\%$) have a younger dynamic age than 4 Myr, 
which is comparable to the lifetime of an O5 star. Most of the bubbles are located
in the spiral arms identified by \citet{Russeil2003}.
The shell density of the bubbles, derived form the Hi-GAL column density maps,
show a decreasing trend with dynamic ages,
suggesting that the bubbles are formed in high-density regions.
We argued that  the shell density tends to decrease with time as a result of the expansion of the shell volume
and is presumably due to the conversion of the shell mass into stars.

We estimate that the CFE of the Hi-GAL clumps located in the bubbles ($< R_{\rm shell}$)
is $\sim10\%$, which is approximately twice higher than the clumps located outside ~ ($R_{\rm shell} < r < 4~R_{\rm eff}$) and is also higher than the CFE reported in other known active star-forming regions. 
We find, however, that the CFE is higher for protostellar than prestellar clumps.
In particular,
the CFE of protostellar clumps has a wider spread 
of values that are higher for younger bubbles and decrease with dynamic ages of the bubbles.
We interpret this trend as a possible increase in the formation rate from the prestellar to protostellar phase
at the early stages of the bubble expansion, which would eventually decrease
as the impact of the expansion and the ionization weakens.

The fraction of clumps that is spatially associated 
with bubbles is $\sim 23\%$, consistent with 
the fraction of ATLASGAL clumps in the
vicinity of MWP bubbles \citep{Kendrew2016}.
However, for the individual fraction of protostellar clumps 
we obtain 41\%. We argue that the higher fraction of protostellar clumps may be related with the
higher protostellar clump formation rate in bubbles.

Bubbles younger than 2~Myr on average have more RMS MYSOs,
and more of them are associated with at least one MYSO 
than in older bubbles. This may be related to the higher density of shells 
in younger bubbles, which is necessary to form MYSOs.
We discussed whether the MYSOs have had time to 
form through gravitational instability of the shell or whether another physical mechanism may be at play.
The shell fragmentation times estimated for this purpose
(following Whitworth et al. 1994\nocite{Whitworth1994}) showed
that a significant 
fraction of bubble shells would not have had time to fragment.
We therefore argued that dense structures
existed in the medium before the bubble expansion to allow for a more comparable star-formation timescale, as 
shown in the simulations performed in \citet{Walch2012,Walch2015}.

The results reported here
point toward the possibility that 
the YSOs we found inside the bubbles
were probably undergoing their formation process before the feedback from the ionizing source(s).
On the other hand, Hi-GAL clumps and RMS MYSOs,
which trace more recent star-formation activity,
indicate that these object were probably triggered
in the shell of the bubbles through a combination of C\&C 
and RDI processes.
Furthermore, the high fraction and CFE of protostellar clumps 
in the early stages of the bubble expansion 
where the density of the shell is higher
might indicate 
an acceleration of the star-formation process through the 
feedback effect of the (H{\sc{ii}}) bubbles.
 

\begin{acknowledgements}
This work is part of the VIALACTEA Project, a Collaborative Project under Framework Programme 7 of the European Union, funded under Contract \#607380, which is hereby acknowledged. \herschel Hi-GAL data processing, map production, and source catalog generation is the result of a multi-year effort that was initially funded thanks to Contracts I/038/080/0 and I/029/12/0 from ASI, Agenzia Spaziale Italiana.

Pedro Palmeirim acknowledges support from the Funda\c{c}\~ao para a Ci\^encia e a Tecnologia of Portugal (FCT)
through national funds (UID/FIS/04434/2013) and by FEDER through COMPETE2020
(POCI-01-0145-FEDER-007672) and also by the fellowship SFRH/BPD/110176/2015 funded by FCT
(Portugal) and POPH/FSE (EC).

\herschel is an ESA space observatory with science instruments provided by European-led Principal Investigator consortia and with important participation from NASA.
This work is based on observations obtained with \herschel-PACS and \herschel-SPIRE photometers. PACS has been developed by a consortium of institutes led by MPE (Germany) and including UVIE (Austria); KU Leuven, CSL, IMEC (Belgium); 
CEA, LAM (France); MPIA (Germany); INAF-IFSI/OAA/OAP/OAT, LENS, SISSA (Italy); IAC (Spain). This development has been supported by the funding agencies BMVIT (Austria), ESA-PRODEX (Belgium), CEA/CNES (France), DLR (Germany), ASI/INAF (Italy), and CICYT/MCYT (Spain). SPIRE has been developed by a consortium of institutes led by Cardiff Univ. (UK) and including: Univ. Lethbridge (Canada);
NAOC (China); CEA, LAM (France); IFSI, Univ. Padua (Italy);
IAC (Spain); Stockholm Observatory (Sweden); Imperial College
London, RAL, UCL-MSSL, UKATC, Univ. Sussex (UK); and Caltech,
JPL, NHSC, Univ. Colorado (USA). This development has been
supported by national funding agencies: CSA (Canada); NAOC
(China); CEA, CNES, CNRS (France); ASI (Italy); MCINN (Spain);
SNSB (Sweden); STFC, UKSA (UK); and NASA (USA). 

This work is based on observations made with the \spitzer Space Telescope, which is operated by the Jet Propulsion Laboratory, California 
Institute of Technology, under contract with NASA. We have made use of the NASA/IPAC Infrared Science Archive to obtain data products from 
the 2MASS, \spitzer-GLIMPSE, and \spitzer-MIPSGAL surveys.
This publication has been made possible by the participation of more than 35,000 volunteers on the Milky Way Project.
Their contributions are acknowledged individually at http://www.milkywayproject.org/authors.

\end{acknowledgements}

\bibliography{Palmeirim_HII_bubbles}

\Online

\end{document}